\documentclass{aa}

\usepackage{times}
\usepackage{graphicx}
\usepackage{xspace}
\usepackage{epsfig}
\usepackage{natbib}
\usepackage{rotating}

\begin{document}

\title{On the origin of the X-ray emission from Herbig Ae/Be stars} 

\author{B. Stelzer\inst {1} \and G. Micela\inst {1} \and K. Hamaguchi\inst {2} \and J. H. M. M. Schmitt\inst {3}}

\offprints{B. Stelzer}

\institute{INAF - Osservatorio Astronomico di Palermo,
% INST 1
  Piazza del Parlamento 1,
  I-90134 Palermo, Italy \\ \email{B. Stelzer, stelzer@astropa.unipa.it} \and
% INST 2
  Exploration of the Universe Division, 
  NASA Goddard Space Flight Center, 
  Greenbelt, MD 20771, USA \and
% INST 3
  Hamburger Sternwarte, 
  Gojenbergsweg 12,
  D-21029 Hamburg, Germany}

\titlerunning{On the origin of the X-ray emission from HAeBe stars}

\date{Received $<$13-02-2006$>$ / Accepted $<$18-05-2006$>$}

\abstract
% context heading (optional)
{Herbig Ae/Be stars are fully radiative and not expected to support dynamo action
analogous to their convective lower-mass counterparts, the T Tauri stars. 
Alternative X-ray production mechanisms,
related to stellar winds or star-disk magnetospheres have been proposed, but all in all 
their X-ray emission has remained a mystery.  
}
% aims heading (mandatory)
{A study of Herbig Ae/Be stars' global X-ray properties (such as detection rate, luminosity, temperature,
variability), helps to constrain the emission mechanism by comparison to other types of
stars, e.g. similar-age but lower-mass T Tauri stars, similar-mass but more evolved
main-sequence A- and B-type stars, and with respect to model predictions.  
}
% methods heading (mandatory)
{We performed a systematic search for Chandra archival observations of Herbig Ae/Be stars. 
The superior spatial resolution of this satellite with
respect to previous X-ray instrumentation has allowed us to examine also the possible role of
late-type companions in generating the observed X-rays. 
}
% results heading (mandatory)
{In the total sample of $17$ Herbig Ae/Be stars, $8$ are resolved from X-ray emitting faint companions
or other unrelated X-ray bright objects within $10^{\prime\prime}$. 
The detection fraction of Herbig Ae/Be stars is $76$\,\%, but lowers to $35$\,\%
if all emission is attributed to further known and unresolved companions. 
The spectral analysis confirms the high X-ray temperatures ($\sim 20$\,MK) and large range of
fractional X-ray luminosities ($\log{L_{\rm x}/L_{\rm *}}$) of this class derived from earlier 
studies of individual objects.
}
% conclusions heading (optional)
{Radiative winds are ruled out as emission mechanism on basis of the high temperatures.
The X-ray properties of Herbig Ae/Be stars are not vastly different from those of their late-type
companion stars (if such are known), nor from other young late-type stars used for comparison. Therefore,
either a similar kind of process takes place on both classes of objects, or
there must be as yet undiscovered companion stars.
}  


\maketitle

\section{Introduction}\label{sect:intro}

Herbig Ae/Be (HAeBe) stars are intermediate-mass stars on the pre-main sequence (pre-MS).
With spectral types A to F they represent high-mass counterparts to
the T Tauri stars (TTS). \citet{Herbig60.1} first introduced HAeBe stars as a class. 
There are no clear criteria for the definition of this group. 
One of their major distinction is strong H$\alpha$ emission, 
but according to some definitions an infrared (IR) excess from dusty circumstellar material
is considered sufficient for attributing a star to the HAeBe class \citep[see][]{Waters98.1}.
In this sense HAeBe stars are closely related to the debris disk systems such as 
$\beta$\,Pic and Vega, which are probably more evolved. 
The most comprehensive catalog of HAeBe stars has been compiled by \citet{The94.1},
including both well-established members of the class and objects that follow a 
more loose definition, such as emission line stars of spectral type F, non-emission line
IR excess stars, and stars that are not found in young associations but appear isolated. 

First systematic studies of X-ray emission from HAeBe stars were presented by 
\citet{Zinnecker94.1} and by \citet{Damiani94.1} using data obtained by the {\em ROSAT} 
and {\em Einstein} missions. 
About $50$\,\% and $30$\,\% of the observed targets were detected, respectively. 
An extensive study using {\em ASCA} and {\em ROSAT} data including spectral 
analysis was published by \citet{Hamaguchi05.1}, and \citet{Skinner04.1} discussed
a small sample with emphasis on an {\em XMM-Newton} observation of HD\,104237. 

The X-ray emission from lower-mass pre-MS stars (TTS) arises from magnetically heated and confined
coronal structures, possibly with a (small) contribution from accretion. 
However, HAeBe stars are on radiative tracks, and therefore not expected to
drive dynamos able to produce the complex field structure that characterizes stars with convective
envelopes. For similar reasons, intermediate-mass stars on the MS are believed to be X-ray dark 
\citep{Schmitt85.1}.
In contrast to the A- and B-type stars on the MS the HAeBe stars may possess a shear dynamo
that is supplied by the rotational energy of the star. This mechanism was shown to sustain magnetic fields
in the initial phase (a few Myrs) of the life of an intermediate-mass star 
\citep{Tout95.1}.
Furthermore, HAeBe stars may be young enough to have maintained their primordial fields. 
However, these fields should be well-ordered and their efficiency in generating X-rays is unclear. 
Direct measurements of magnetic fields on HAeBe stars are
difficult because of the presence of circumstellar matter, the large number of emission lines,
and their generally fast rotation. A few marginal detections were reported by \citet{Donati97.1}, 
\citet{Hubrig04.1}, and \citet{Wade05.1}. In the latter study $5$ of $50$ HAeBe stars observed
with spectropolarimetric methods showed magnetic signatures in the Stokes $V$ profile,
indicating longitudinal fields of a few hundred Gauss at most. 

The detection of non-thermal radio emission would be evidence for the presence of magnetic
fields, but most of the radio detections among the HAeBe stars seem to be thermal sources,
and were interpreted as being due to a weak stellar wind \citep{Skinner93.1}. 
Further evidence for winds comes from forbidden line emission \citep{Corcoran97.1}.
Winds can also explain the various observed shapes of the H$\alpha$ emission profile in HAeBe stars
\citep{Boehm95.1}. 
Instabilities in the wind could give rise to shocks that heat the plasma to X-ray emitting
temperatures, in analogy to the case of OB stars. 
The early X-ray studies of HAeBe stars with {\em Einstein} and {\em ROSAT} 
\citep{Damiani94.1, Zinnecker94.1} came up with
correlations between X-ray luminosity and kinetic wind energy ($\dot{M}v_\infty^2$) as well 
as wind loss rate ($\dot{M}$). However, subsequent spectral studies with {\em ASCA} showed X-ray
temperatures in excess of $1$\,keV \citep{Hamaguchi05.1}. The winds of HAeBe stars are weaker than those of hot stars, 
and their terminal speed is probably not sufficient to produce these temperatures \citep{MacFarlane89.1}. 
Magnetic fields -- if present -- may influence the wind geometry and dynamics by channeling the wind 
as in the MCWS ( = Magnetically Confined Wind Shock) model. 
This scenario was originally developed by \citet{Babel97.1} 
to explain the X-ray emission from the Ap star IQ\,Aur, but was also applied to the O star
$\Theta$\,Ori\,C \citep{Babel97.2}. The same mechanism could be the cause for the strong and 
variable X-ray emission of some hot stars in Orion \citep{Stelzer05.1}.
Its application to HAeBe stars depends on whether wind velocities and X-ray temperatures
can be reconciled, and seems to be dubious according to previous results. 

Some HAeBe stars are still in the accretion phase.  
The possible role of accretion for the generation of X-rays in low-mass pre-MS stars
has recently gained support, 
most importantly by the high density and soft spectrum of the emitting plasma 
derived from high-resolution X-ray spectroscopy \citep{Kastner02.1, Stelzer04.2, Schmitt05.1}. 
Equivalent studies for HAeBe stars are absent from the literature. 
A possible link between X-ray activity of HAeBe stars and the presence of outflows, 
typical for stars in the accretion phase, was suggested by \citet{Hamaguchi05.1}. 

Despite the abundant speculations about the possible mechanism, 
a convincing and unique explanation for the X-ray emission of HAeBe stars has not been identified. 
A further hypothesis is that of unknown/unresolved T Tauri like companions,
that would be responsible for the X-ray emission. This idea is the most favored explanation for
the observed X-rays from more evolved A- and B-type stars on the MS. \citet{Stelzer03.1} and 
\citet{Stelzer06.1} have checked the companion hypothesis by resolving a sample of B-type
stars with {\em Chandra} from their close visual companions. The result was ambiguous, because
more than half of the B-type stars were detected with {\em Chandra} even after being
resolved from all known visual companions. However, the X-ray sources associated with the B-type
stars do not show significantly different properties from those coincident with their late-type
companion stars. This is fully consistent with the idea that the X-rays originate from 
even closer spectroscopic companions. 

Here we apply the same approach to the sample of HAeBe stars observed with {\em Chandra}.
{\em Chandra} is the only satellite that provides a sub-arcsecond spatial resolution in X-rays,
which is reasonably close to IR imaging (adaptive optics) observations. 
This implies that the majority of known visual companions,
those at separations larger than $\sim 1^{\prime\prime}$, can be resolved for the first time in X-rays. 

Our knowledge on multiplicity of HAeBe stars is rather incomplete. 
A few studies have focused on a search for close visual binaries \citep[e.g. ][]{Li94.1,
Pirzkal97.1, Leinert97.1, Smith05.1}. 
Monte Carlo simulations by \citet{Pirzkal97.1}
have suggested that almost all HAeBe stars have companions within a completeness limit of
$K < 10.5$\,mag for separations of $0.4...8^{\prime\prime}$.  
\citet{Corporon99.1} performed a spectroscopic survey for binaries among HAeBe stars,
and discovered $13$ binaries by means of radial velocity variations or the detection of  
Lithium absorption. 

The paper is structured as follows. 
In Sect.~\ref{sect:sample} we introduce the targets 
observed with {\em Chandra}.  The observations and the data analysis are 
described in Sect.~\ref{sect:observations}.  
Sect.~\ref{sect:results} summarizes the results from source detection,
the temporal and the spectral analysis. In the discussion (Sect.~\ref{sect:discussion})
we examine the nature of the X-ray emitters, the global X-ray properties of HAeBe stars, 
the $L_{\rm x}/L_{\rm bol}$ relation,
and correlations between X-ray emission and other characteristic stellar parameters. 
Sect.~\ref{sect:conclusions} presents the conclusions. Information on individual
stars is found in the Appendix~\ref{sect:indiv}.

\section{The Sample}\label{sect:sample}

\begin{figure}
\begin{center}
\resizebox{9cm}{!}{\includegraphics{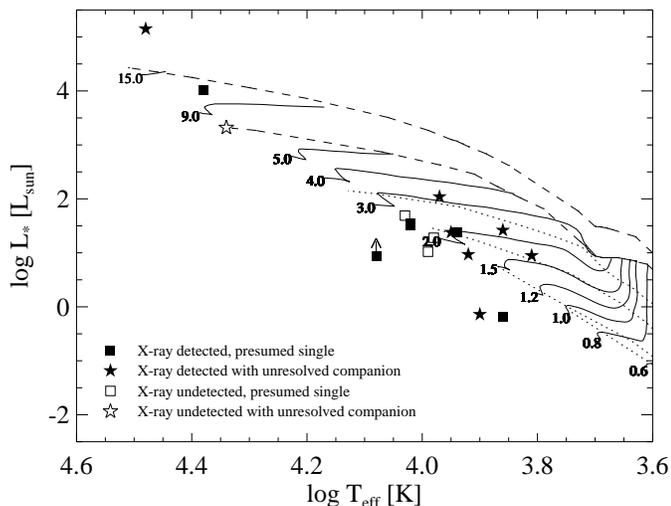}}
\caption{HR diagram for HAeBe stars observed with {\em Chandra}. 
Superposed on the data are evolutionary models by \protect\citet{Palla99.1}: solid lines are tracks labeled
by the corresponding mass in solar units, dotted lines are $1$, $5$, and $30$\,Myr isochrones and dashed lines
are the birthlines for accretion rates of $\dot{M} = 10^{-4}\,{\rm M_\odot/yr}$ and
$10^{-5}\,{\rm M_\odot/yr}$, respectively.}
\label{fig:hrd}
\end{center}
\end{figure}

We have searched the {\em Chandra} archive for observations of HAeBe stars.   
Positional cross-correlation with the catalogs by \citet{The94.1},  
\citet{vandenAncker97.1}, and \citet{vandenAncker98.1} turned up 
$17$ objects. Their position in the HR diagram is shown in Fig.~\ref{fig:hrd}. 
The stellar parameters were compiled from the literature
and are given together with the references in Table~\ref{tab:early_hrd}. 
Two stars, V892\,Tau and R\,CrA, are located well below the MS in Fig.~\ref{fig:hrd}, 
probably indicating a serious problem in the calculation of $L_{\rm *}$ due to the 
strong and poorly constrained extinction; see also Sect.~\ref{subsect:xray_lbol}.
%
%  OUTPUT FROM    plot_tracks.pro
%
\begin{table*}
\begin{center}
\caption{Stellar parameters for HAeBe stars observed with {\em Chandra} (left side) and parameters relevant for the interpretation of the X-ray emission (right side). The references in col.~5 refer to cols.~$2-4$, separated by commas. The flags in cols.~$8-10$ stand for X-ray detection (`X'), presence of unresolved visual companion (`VB') and spectroscopic companion (`SB').} 
\label{tab:early_hrd}
\begin{tabular}{lrrrcrr|ccc}\hline
Designation & $A_{\rm V}$ & $\log{T_{\rm eff}}$ & $\log{(L_{\rm bol}/L_\odot)}$ & Ref & $v \sin{i}$  & Ref & X & VB & SB \\ 
            & [mag]       & [K]                &                              &     & [km/s]         &     &   &    &    \\ 
\hline
BD$+30^\circ549$ & 1.89       & 4.08       & $>$0.93        & (1,1,1)     &        &     & $\surd$ &         &         \\
V892\,Tau        & 3.9...12.5 & 3.90       & -0.14          & (7,8,8)     &        &     & $\surd$ & $\surd$ &         \\
V380\,Ori        & 1.43...1.7 & 3.97       & 2.04           & (1...2,1,3) & 200    & (4) & $\surd$ & $\surd$ & $\surd$ \\
HD\,147889       & 3.32       & 4.34       & 3.32           & (6,6,6)     & 30..70 & (4) & $-$     &         & $\surd$ \\
V590\,Mon        & 0.6...1.3  & 3.99       & 1.02           & (2...5,2,3) &        &     & $-$     &         &         \\
Z\,CMa           & 2.4...4.46 & 3.8...4.48 & $>1.72$...5.15 & (7...8,1...8,1...8) & $<130$ & (1) & $\surd$ & $\surd$ & \\
HD\,97300        & 1.33       & 4.02       & 1.54           & (6,6,6)     &        &     & $\surd$ &         &         \\
HD\,100546       & 0.2...1.0  & 4.02       & 1.51           & (7,1,1)     & 250,55 & (1,9)&$\surd$ &         &         \\
HD\,104237       & 0.3...1.3  & 3.86       & 1.42           & (7,8,8)     & 10     & (9) & $\surd$ &         & $\surd$ \\
HD\,141569       & 0.47       & 3.98       & 1.28           & (1,8,8)     & 236,200& (1,4)&$-$     &         &         \\
HD\,150193       & 1.61       & 3.95       & 1.38           & (1,8,8)     & 100    & (8) & $\surd$ &         & $\surd$ \\
HD\,152404       & 0.62       & 3.81       & 0.88...0.95    & (8,1,1...8) & 19     & (1) & $\surd$ &         & $\surd$ \\
HD\,163296       & 0.09...0.3 & 3.94       & 1.38...1.48    & (8...7,8,8...1) & 120& (10)& $\surd$ &         &         \\
MWC\,297         & 2.9...8.3  & 4.38       & 4.01           & (7,8,8)     &    &         & $\surd$ &         &         \\
HD\,176386       & 0.62       & 4.03       & 1.69           & (6,6,6)     &        &     & $-$     &         &         \\
TY\,CrA          & 1.0...3.1  & 3.92       & 0.97           & (7,2,2)     & 10,88  &(4,2)& $\surd$ & $\surd$ & $\surd$ \\
R\,CrA           & 0.9...1.9  & 3.86...3.81& -0.19...-0.38  & (7,8...2,8...2) &    &     & $\surd$ &         &         \\
\hline
\multicolumn{10}{l}{(1) - \protect\citet{vandenAncker98.1}, (2) - \protect\citet{Hillenbrand92.1}, (3) - \protect\citet{Corcoran98.1}, (4) - \protect\citet{Corporon99.1},} \\
\multicolumn{10}{l}{(5) - \protect\citet{Hernandez04.1}, (6) - \protect\citet{vandenAncker97.1}, (4) - \protect\citet{Haffner95.1}, (7) - \protect\citet{Elia04.1},} \\
\multicolumn{10}{l}{(8) - \protect\citet{Acke04.1}, (9) - \protect\citet{Acke04.2}, (10) - \protect\citet{Finkenzeller85.1}.} \\
\end{tabular}
\end{center}
\end{table*}

In Fig.~\ref{fig:hrd} and in Table~\ref{tab:early_hrd} we anticipate the 
results from the {\em Chandra} source detection process discussed further 
in Sect~\ref{subsect:srcdet}. 
The last three columns of Table~\ref{tab:early_hrd} summarize the X-ray detection status of 
all HAeBe stars, and whether there are known companions that remained unresolved with {\em Chandra}. 
In Fig.~\ref{fig:hrd} binaries that remain unresolved with {\em Chandra} 
are marked with different plotting symbols. 
Information on companions was collected from the literature.  
We focus on close systems, that could not be resolved with previous X-ray satellites.   
About half of these companions are separable from the HAeBe stars with {\em Chandra} 
(those with separation $\geq 1^{\prime\prime}$). 
We point out that physical companionship has not been established for most of these objects. 
Confirmation that they are true companions, and not just chance projections, 
requires observations of their proper motion and/or spectra. 

For many of the targets the {\em Chandra} data 
have already been discussed in some detail in the literature, 
but no homogeneous analysis of these observations has been presented so far. 
Furthermore, we add some updates and corrections with respect to
previous publications. 
The list of all targets is summarized in Table~\ref{tab:obslog} that gives 
the designation, position, distance, and spectral type of the HAeBe star, 
as well as separation, position angle, an identifier flag for all known companions, 
and the respective reference. 
Some basic information about the {\em Chandra} observations is given at the end
of Table~\ref{tab:obslog}, where the reference column refers to previous citations of the
{\em Chandra} data. 

The data were obtained within various unrelated observing programs. Therefore, 
the sample observed with {\em Chandra} may not be representative of the total population
of HAeBe stars. As can be seen from Fig.~\ref{fig:hrd}
the stars are distributed unevenly across the HR diagram. The observations are also not 
homogeneous in terms of sensitivity limit, because of the different distances and exposure times. 
Most important for our study is a possible bias towards known X-ray emitters. 

%
% Position + distance of primary from Hipparcos data base
% Sptype from Berghoefer et al (1996)
% Separation and Position angle from Huelamo et al (2001), Hubrig et al. (2001),
%   and Shatsky & Tokovinin (2002)
% 
\begin{table*}\tiny
\begin{center}
\caption{Target list of HAeBe stars observed with {\em Chandra}: position, distance, and spectral type of primaries; separation, position angle, and identifier of companion candidates; observation ID, instrument, exposure time, and reference to previous publication for {\em Chandra} observations.} 
\label{tab:obslog}
\begin{tabular}{lrrrlrrccrcrr}
\noalign{\smallskip} \hline \noalign{\smallskip} 
\multicolumn{5}{c}{HAeBe primaries} & \multicolumn{4}{c}{Companions} & \multicolumn{4}{c}{ACIS observations} \\
Designation & \multicolumn{2}{c}{Position$^{(a)}$}                                       & Dist$^{(a)}$ & SpT$^{(b)}$ & \multicolumn{1}{c}{Sep$^{(c)}$}                   & \multicolumn{1}{c}{PA$^{(c)}$}         & Component & Ref. & ObsID & Instr & Expo  & Ref. \\
            & \multicolumn{1}{c}{$\alpha_{2000}$} & \multicolumn{1}{c}{$\delta_{2000}$}  & [pc]          &             & \multicolumn{1}{c}{[$^{\prime\prime}$]}  & \multicolumn{1}{c}{[$^\circ$]} &       &      &  &  & [s] &  \\
\noalign{\smallskip} \hline \noalign{\smallskip} 
BD+30$^\circ$549  & 03:29:19.78 & $+$31:24:57.0 & 390     & B8\,V\,pe    & $-$    & $-$           &   &     & 0642 & I & 43351.2 & (17) \\
\noalign{\smallskip} \hline \noalign{\smallskip}
V892\,Tau   & 04:18:40.61 & $+$28:19:15.5$^{*}$ & 162$^*$ & A6\,e        & $4.10$ & $ 23.4$       & B & (1) & 3364 & S & 17734.2 & (18) \\
            &             &                     &         &              & $0.05$ & $ 50$         & C & (2) &      &   &         &      \\
\noalign{\smallskip} \hline \noalign{\smallskip} 
V380\,Ori   & 05:36:25.43 & $-$06:42:57.7       & 460$^*$ & B8/A1\,e     & $0.15$ & $204.2$       & B & (1) & 0021 & S & 19510   &      \\
            &             &                     &         &              & \multicolumn{2}{c}{SB} &   & (3) &      &   &         &      \\
\noalign{\smallskip} \hline \noalign{\smallskip}
HD\,147889  & 06:25:24.32 & $-$24:27:56.6       & 136     & B2\,V        & \multicolumn{2}{c}{SB} &   & (4) & 0618 & I &  5014.1 &      \\
\noalign{\smallskip} \hline \noalign{\smallskip} 
V590\,Mon   & 06:40:44.64 & $+$09:48:02.2$^{*}$ & 800$^*$ & B8\,ep+sh    & $-$    & $-$           &   &     & 2550 & I & 48138   & (19) \\
\noalign{\smallskip} \hline \noalign{\smallskip} 
Z\,CMa      & 07:03:43.16 & $-$11:33:06.2       & 1050$^*$& F6\,III\,e   & $0.11$ & $129$         & B & (5) & 3751 & S &         &      \\
\noalign{\smallskip} \hline \noalign{\smallskip}
HD\,97300   & 11:09:50.02 & $-$76:36:47.7       & 188     & B9\,V        & $-$    & $-$           &   &     & 1867 & I & 66292.1 & (20) \\
\noalign{\smallskip} \hline \noalign{\smallskip} 
HD\,100546  & 11:33:25.44 & $-$70:11:41.2       & 103     & B9\,V\,ne    & $4.54$ & $196.5$       & & (6) & 2403,3427 & I & 2622,2628 & (21) \\
            &             &                     &         &              & $5.22$ & $155.1$       & & (6) &           &   &           &      \\
            &             &                     &         &              & $5.91$ & $ 26.4$       & & (6) &           &   &           &      \\
            &             &                     &         &              & $5.55$ & $322.6$       & & (6) &           &   &           &      \\
\noalign{\smallskip} \hline \noalign{\smallskip} 
HD\,104237  & 12:00:05.08 & $-$78:11:34.6       & 116     & A0\,V\,pe    & $5.28$ & $310.1$       & 1 & (7) & 2404,3428 & I & 2904,2828 & (21) \\
            &             &                     &         &              &$ 1.37$ & $254.6$       & 2 & (7) &      &   &         &     \\
            &             &                     &         &              &$10.72$ & $117$         & 5 & (7*) &      &   &         &     \\
            &             &                     &         &              &$14.88$ & $121$         & 6 & (7*) &      &   &         &     \\
            &             &                     &         &              & \multicolumn{2}{c}{SB} &   & (8) &      &   &         &     \\
\noalign{\smallskip} \hline \noalign{\smallskip} 
HD\,141569  & 15:49:57.75 & $-$03:55:16.4       &  99     & B9.5\,V\,e   & $7.57$ & $311.5$       & B & (9) & 0981 & I & 2916    & (21) \\
            &             &                     &         &              & $8.93$ & $310.0$       & C & (9) &      &   &         &      \\
\noalign{\smallskip} \hline \noalign{\smallskip}
HD\,150193  & 16:40:17.92 & $-$23:53:45.2       & 150     & A1\,V\,e     & $1.10$ & $225$         & B &(10) & 0982 & I & 2919    & (21) \\
            &             &                     &         &              & \multicolumn{2}{c}{SB} &   & (3) &      &   &         &      \\
\noalign{\smallskip} \hline \noalign{\smallskip} 
HD\,152404  & 16:54:44.85 & $-$36:53:18.6       & 145     & F5\,V        & \multicolumn{2}{c}{SB} &   &(11) & 0983 & I & 3112.1  & (21) \\
\noalign{\smallskip} \hline \noalign{\smallskip}
HD\,163296  & 17:56:21.29 & $-$21:57:21.9       & 122     & A1\,V\,e     & $-$    & $-$          &   &(12) & 3733 & S & 20000   & (22) \\
\noalign{\smallskip} \hline \noalign{\smallskip}
MWC\,297    & 18:27:39.65 & $-$03:49:52.0$^{*}$ & 250$^*$ & O9\,e        & $3.39$ & $313$         & B &(13) & 1883 & I & 40000   & (15,23)\\
\noalign{\smallskip} \hline \noalign{\smallskip} 
HD\,176386  & 19:01:38.93 & $-$36:53:26.5       & 136     & B9\,IV       & $4.1$  & $138$         & B &(14) & 0019 & I & 37330   &      \\
\noalign{\smallskip} \hline \noalign{\smallskip} 
TY\,CrA     & 19:01:40.83 & $-$36:52:33.9$^{*}$ & 136 ?   & B9\,e        & $0.29$ & $188.5$       & B &(15) & 0019 & I & 37330   & \\
            &             &                     &         &              & \multicolumn{2}{c}{SB} & & (16) &&   &         & \\
\noalign{\smallskip} \hline \noalign{\smallskip} 
R\,CrA      & 19:01:53.65 & $-$36:57:07.6       & 130$^*$ & A1\,e-F7\,e\,var & $-$    & $-$           &   &     & 0019 & I & 37330   & (24) \\
\noalign{\smallskip} \hline \noalign{\smallskip} 
\multicolumn{13}{l}{$*$ For two companions to HD\,104237 the position angle from the literature does not coincide with the published IR image.} \\ % but a visual comparison of the IR and X-ray images allows us} \\
\multicolumn{13}{l}{The position angles cited in this table and used throughout this paper are corrected values (Grady, priv. comm.).} \\ 
\multicolumn{13}{l}{$^{(a)}$ {\em Hipparcos} position and distance; exceptions are data marked with $^*$ where positions are from \citet{Ducourant05.1}, \citet{Hog98.1},} \\
\multicolumn{13}{l}{and \citet{The94.1}, and distances from \citet{Hillenbrand92.1}, \citet{Shevchenko99.1}, \citet{vandenAncker98.1}, and \citet{Drew97.1}.} \\
\multicolumn{13}{l}{$^{(b)}$ spectral types adopted from \protect\citet{The94.1}, \protect\citet{vandenAncker97.1}, or \protect\citet{vandenAncker98.1};} \\
\multicolumn{13}{l}{$^{(c)}$ `SB' stands for spectroscopic companions.} \\
\multicolumn{13}{l}{References:} \\
\multicolumn{13}{l}{(1) - \protect\citet{Leinert97.1}, (2) - \protect\citet{Smith05.1}, (3) - \protect\citet{Corporon99.1}, (4) - \protect\citet{Haffner95.1},} \\
\multicolumn{13}{l}{(5) - \protect\citet{MillanGabet02.1}, (6) - \protect\citet{Shatsky02.1}, (7) - \protect\citet{Grady04.1}, (8) - \protect\citet{Boehm04.1},} \\
\multicolumn{13}{l}{(9) - \protect\citet{Weinberger99.1}, (10) - \protect\citet{Fukagawa03.1}, (11) - \protect\citet{Andersen89.1}, (12) - \protect\citet{Grady00.1}, (13) - \protect\citet{Vink05.1},} \\
\multicolumn{13}{l}{(14) - \protect\citet{Turon93.1}, (15) - \protect\citet{Chauvin03.1}, (16) - \protect\citet{Casey95.1},} \\
\multicolumn{13}{l}{(17) - \protect\citet{Getman02.1}, (18) - \protect\citet{Giardino04.1}, (19) - \protect\citet{Ramirez04.1}, (20) - \protect\citet{Feigelson04.1},} \\
\multicolumn{13}{l}{(21) - \protect\citet{Feigelson03.1}, (22) - \protect\citet{Swartz05.1}, (23) - \protect\citet{Damiani06.1}, (24) - \protect\citet{Forbrich06.1}.} \\ 
\end{tabular}
\end{center}
\end{table*}

\section{Observations and Data Analysis}\label{sect:observations}

Most targets were observed with the standard imaging array of {\em Chandra}'s 
Advanced CCD Imaging Spectrometer (ACIS-I); see \citet{Weisskopf02.1} for
details on the satellite and its instruments. 
In some cases the target was placed on the backside illuminated ACIS-S3 chip. 
The observing instrument is indicated in column~11 
of Table~\ref{tab:obslog}. 

Data analysis was carried out using the CIAO software 
package\footnote{CIAO is made available by the CXC and can be downloaded 
from \\ http://cxc.harvard.edu/ciao/download-ciao-reg.html} version 3.2
in combination with the calibration database (CALDB) version 3.0.0.
We started our analysis with the level\,1 events file provided by the
{\em Chandra} X-ray Center (CXC). 
On all observations processed at the CXC with CALDB version earlier than 2.9 
we applied a new gain map and updates 
on the geometry (focal length, ACIS pixel size and chip positions). For 
observations processed with a later version of CALDB 
these modifications had been performed automatically during the pipeline process.  
In the process of converting the level\,1 events file to a level\,2 events file
for each of the observations we performed the following steps: 
A correction for the charge transfer inefficiency (CTI) has been applied for data
with processing version earlier than 6.12, i.e. for those where the CTI correction
was not yet part of the standard pipeline processing at the CXC. 
We removed the pixel randomization which is automatically applied by the CXC pipeline
in order to optimize the spatial resolution. 
We filtered the events file for event grades
(retaining the standard grades $0$, $2$, $3$, $4$, and $6$), 
and applied the standard good time interval file. 
Events flagged as cosmic rays were not removed in our analysis. In principle, such events 
can lead to the detection of spurious sources. However, if identified on the position of a
bright X-ray source, the flag is often erroneous (as a result of the event pattern used for 
the identification of cosmic rays). 
Inspection of our images revealed that a substantial number of events at the positions of 
our sources carry the cosmic ray flag 
(typically $3-5$\,\%; see http://cxc.harvard.edu/ciao/threads/acisdetectafterglow). 
Removing these events would therefore result in an underestimate of the source count rate. 

Since the positional accuracy is particularly
important to our observations we also checked the astrometry for any known 
systematic aspect offsets using the CIAO aspect 
calculator\footnote{see http://asc.harvard.edu/ciao/threads/arcsec\_correction }. 
Small offsets are present in all of the presented observations,  
and the aspect was corrected accordingly by modifying the respective header keywords 
in the level\,2 events file.

\subsection{Source Detection}\label{subsect:srcdet}

Source detection was restricted to a $50 \times 50$ pixels wide image 
(1\,pixel $= 0.492^{\prime\prime}$), centered on the position of the primary HAeBe star. 
Exceptions are HD\,104237, where the image
center was offset to include all obvious nearby X-ray sources, and R\,CrA where visual 
inspection of the image identifies two X-ray emitters at slightly wider separations, 
and source detection was performed on a $100 \times 100$ pixels large area. 
Source detection was carried out with the {\sc wavdetect} algorithm \citep{Freeman02.1}.
This algorithm correlates the data with a mexican hat function
to search for deviations from the background. The {\sc wavdetect} 
mechanism is well suited for separating closely spaced point sources.  
We used wavelet scales between $1$ and $8$ in steps of $\sqrt{2}$. 
Two of the targets (HD\,100546 and HD\,104237) were observed twice, 
and we merged the observations to obtain a combined image with higher S/N.

\section{Results}\label{sect:results}

Fig.~\ref{fig:acis_images_haebe} shows a portion of the ACIS images 
centered on the HAeBe stars. In general, the scale of the images displayed in this figure
corresponds to the area in which source detection was performed. 
However, in some cases we show a smaller image fraction in order to provide a better view of crowded fields. 
The photon extraction areas of all detected X-ray sources are overplotted (circles), 
as well as the position of the primary and the position of the companions 
(x-shaped symbols). We point out that for the case of HD\,141569 and HD\,150193 our identifications
differ from the results presented by \citet{Feigelson03.1}; some of the erroneous results 
were also used by \citet{Skinner04.1}. For a detailed discussion of this issue we
refer to appendix~\ref{subsect:hd141569} and~\ref{subsect:hd150193}.  
\begin{figure*}
\begin{center}
\parbox{18cm}{
\parbox{6cm}{
\resizebox{6cm}{!}{\includegraphics{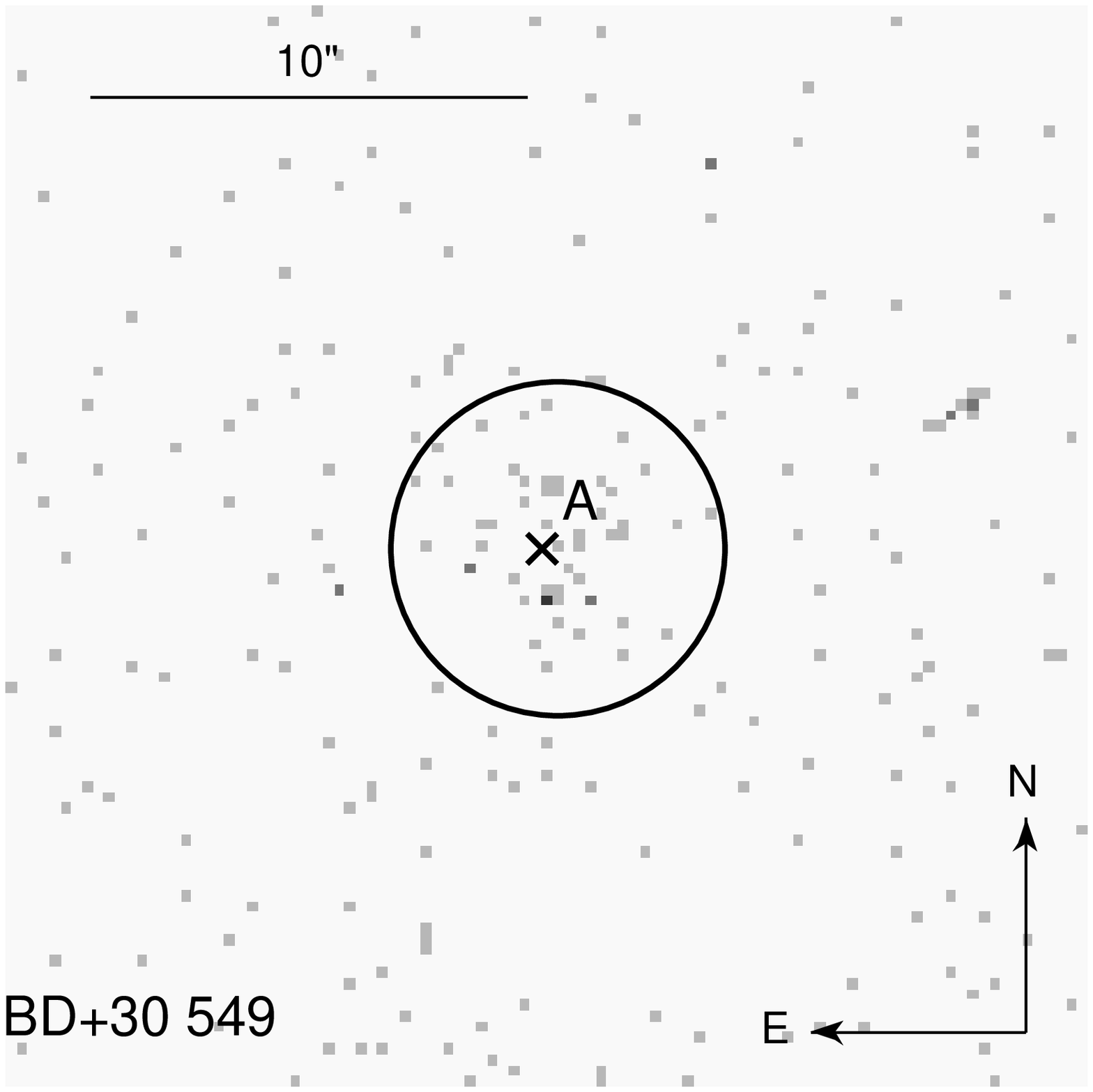}}
}
\parbox{6cm}{
\resizebox{6cm}{!}{\includegraphics{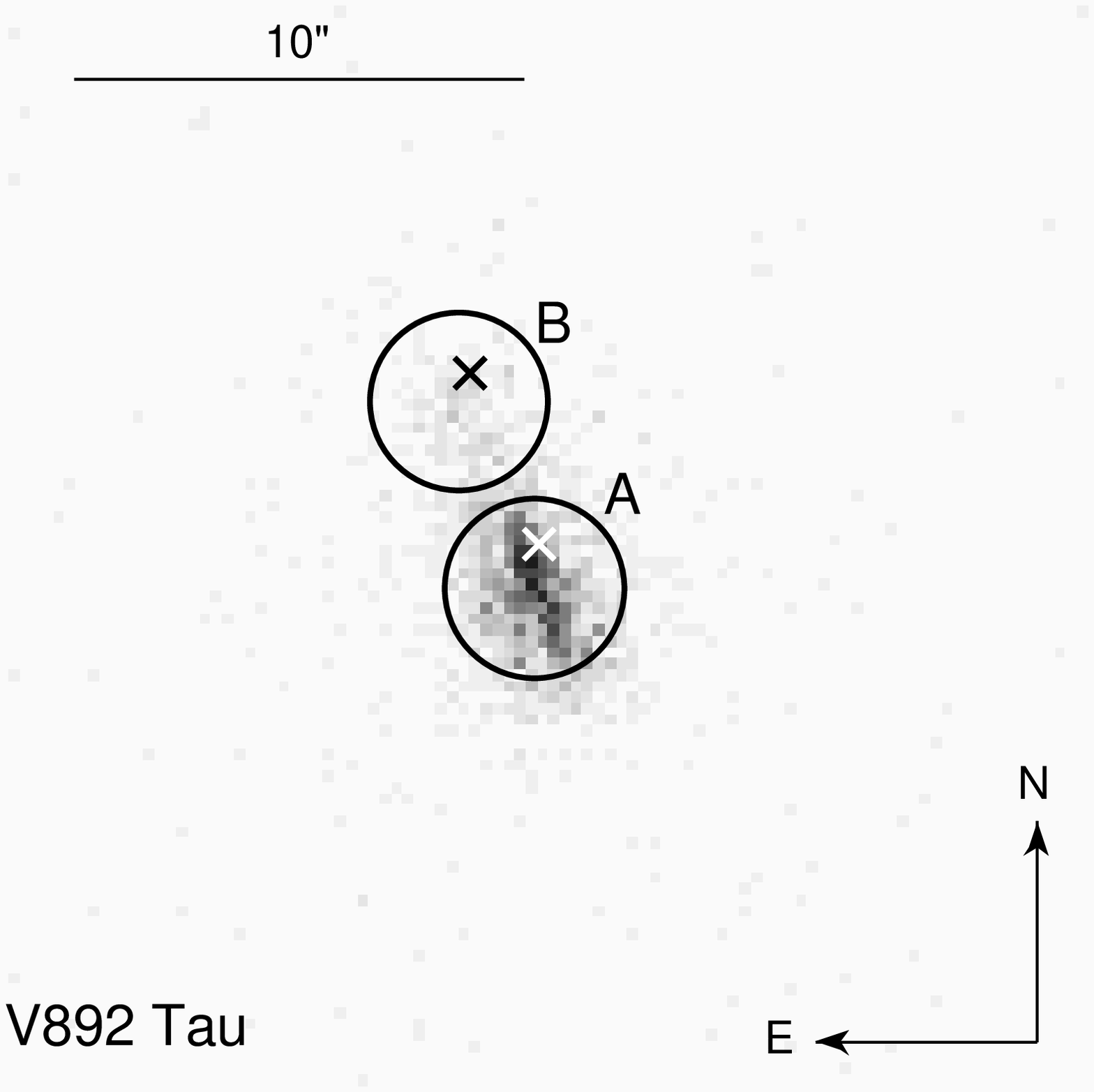}}
}
\parbox{6cm}{
\resizebox{6cm}{!}{\includegraphics{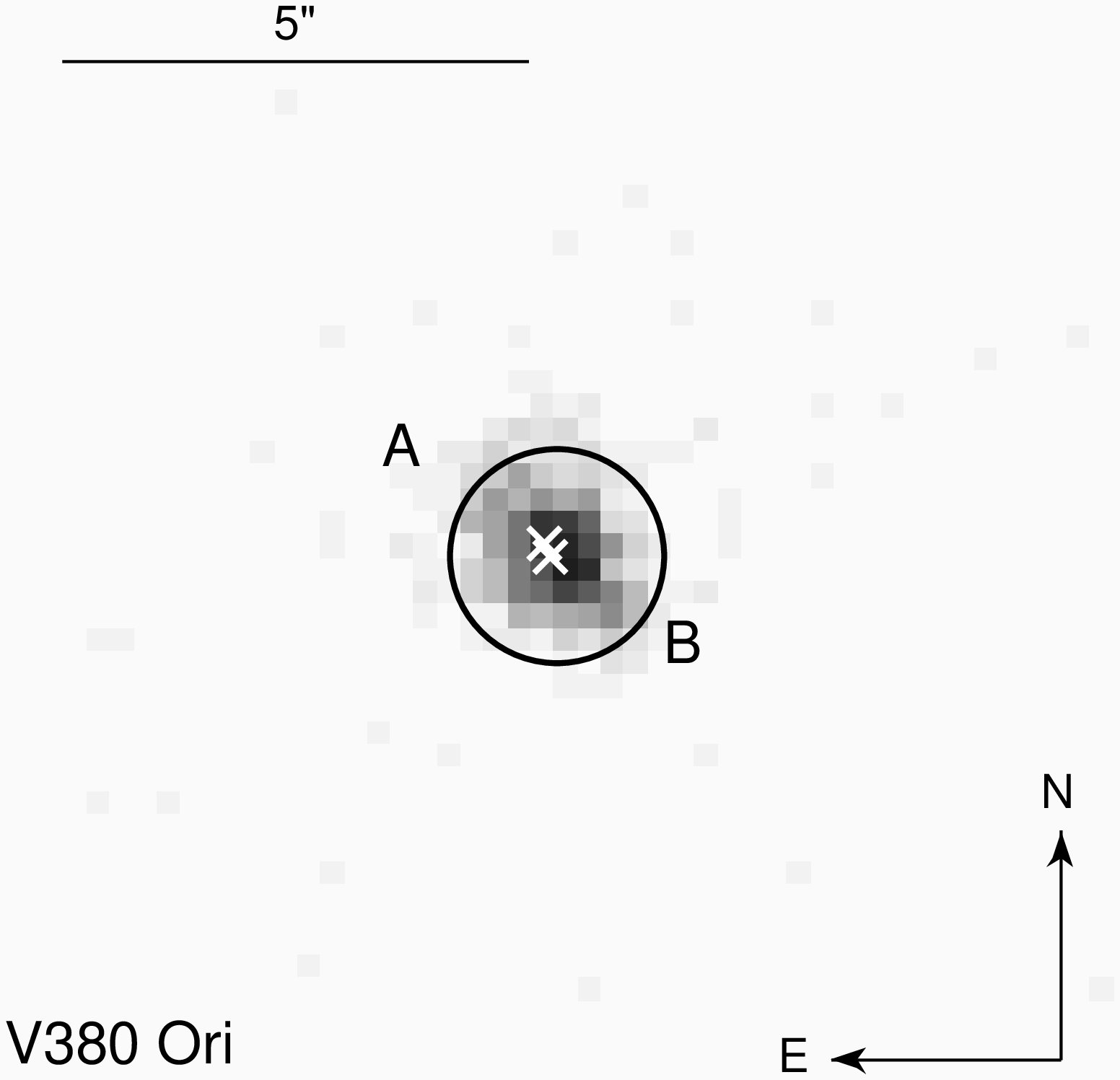}}
}
}
\parbox{18cm}{
\parbox{6cm}{
\resizebox{6cm}{!}{\includegraphics{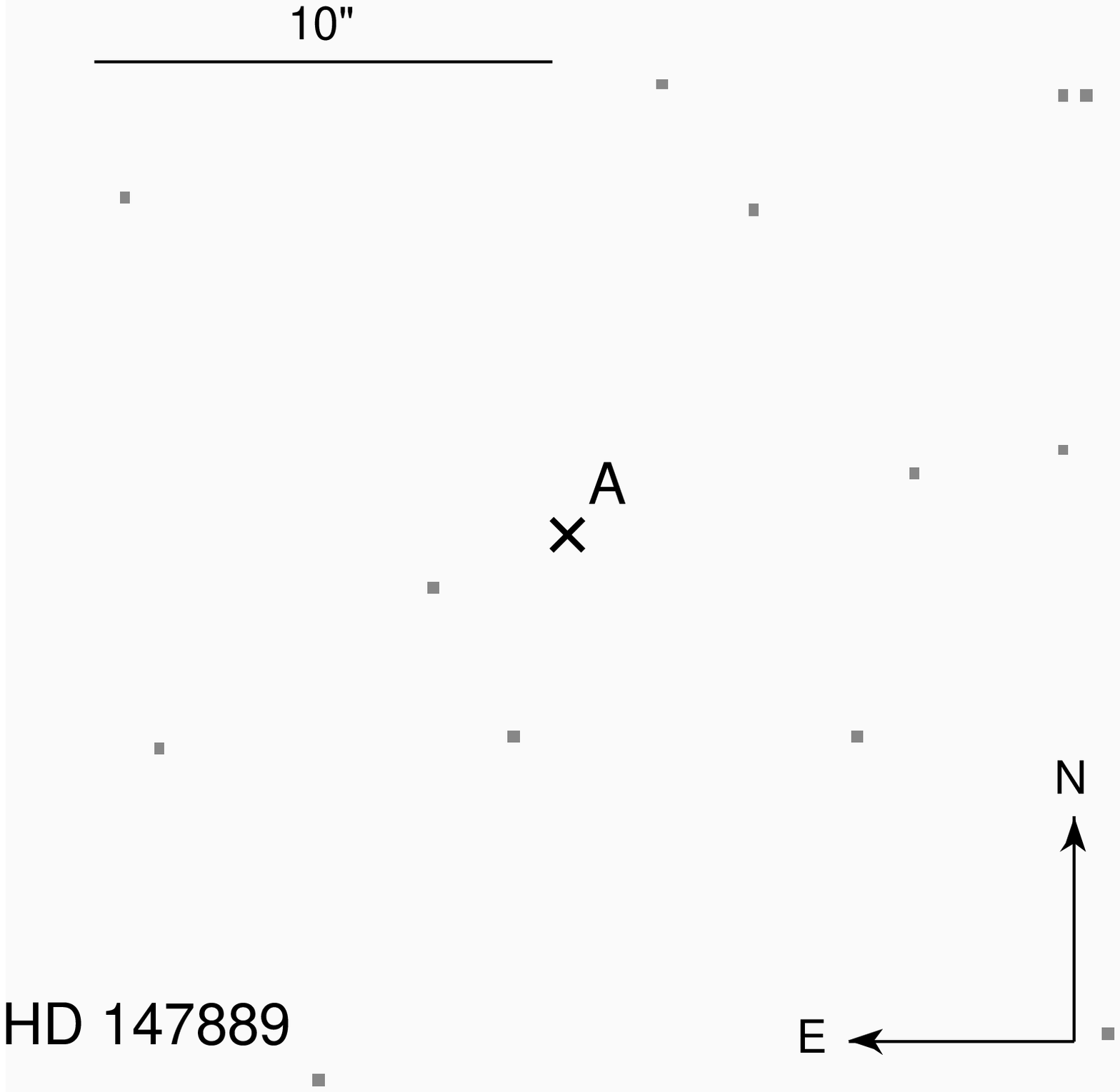}}
}
\parbox{6cm}{
\resizebox{6cm}{!}{\includegraphics{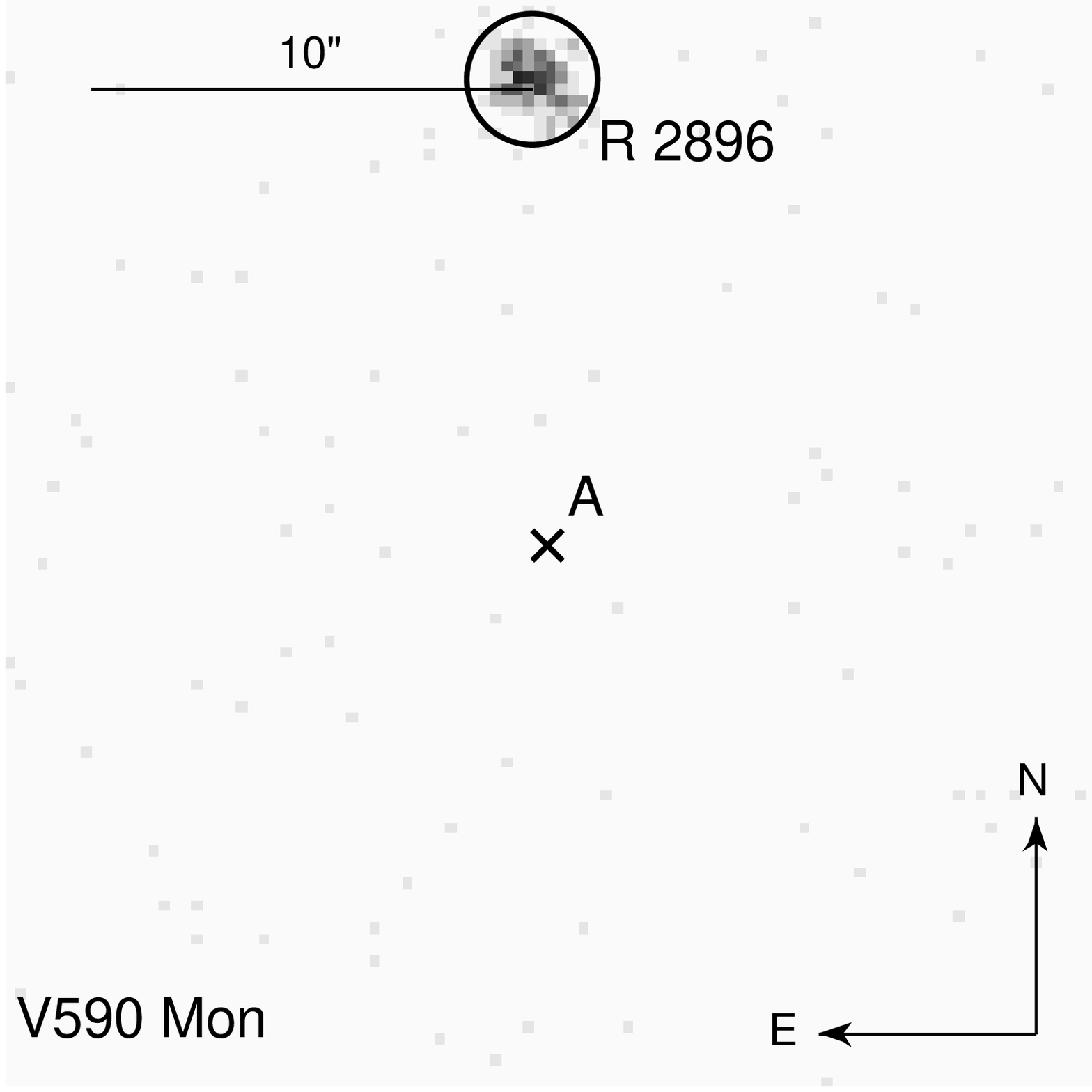}}
}
\parbox{6cm}{
\resizebox{6cm}{!}{\includegraphics{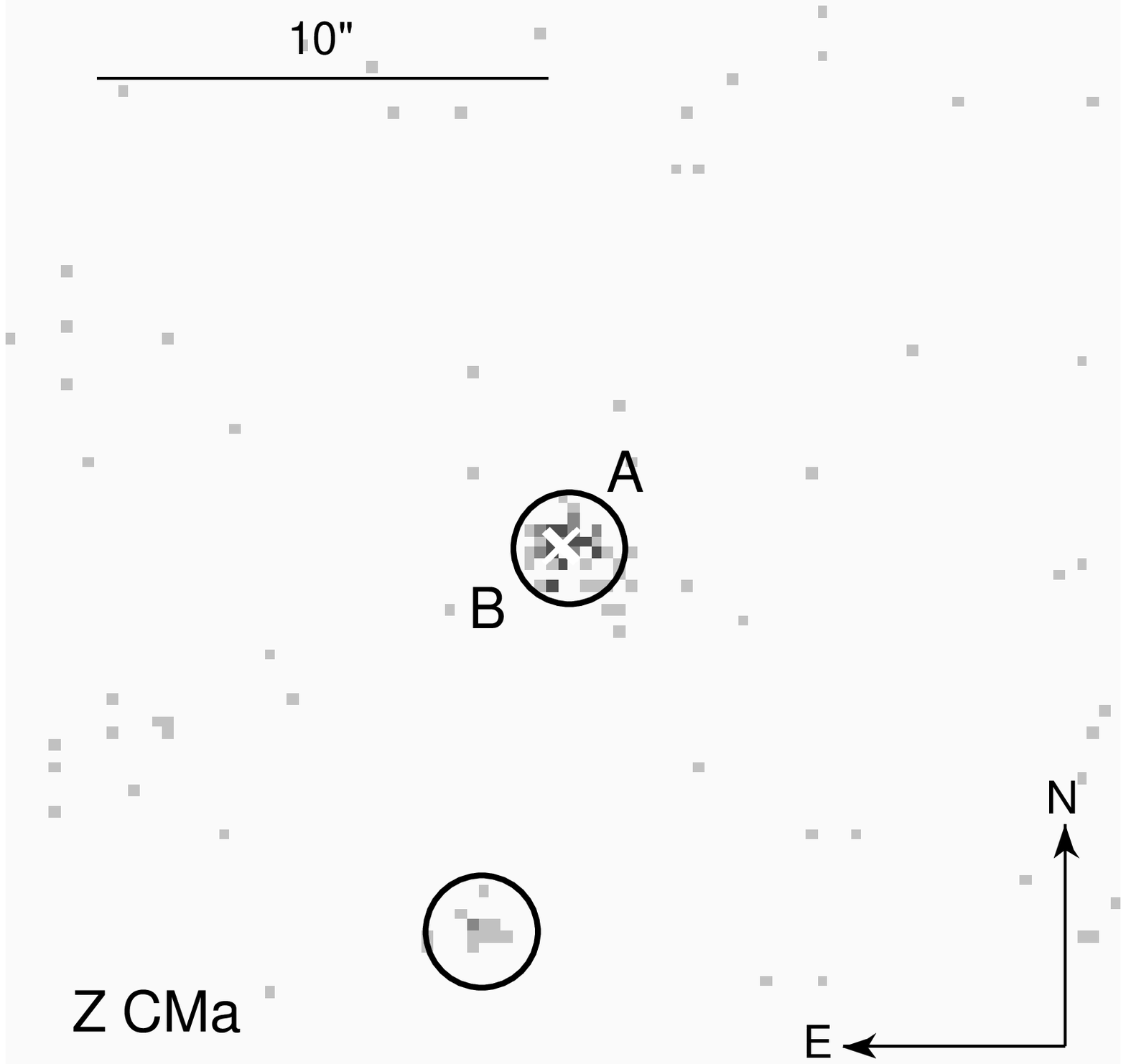}}
}
}
\parbox{18cm}{
\parbox{6cm}{
\resizebox{6cm}{!}{\includegraphics{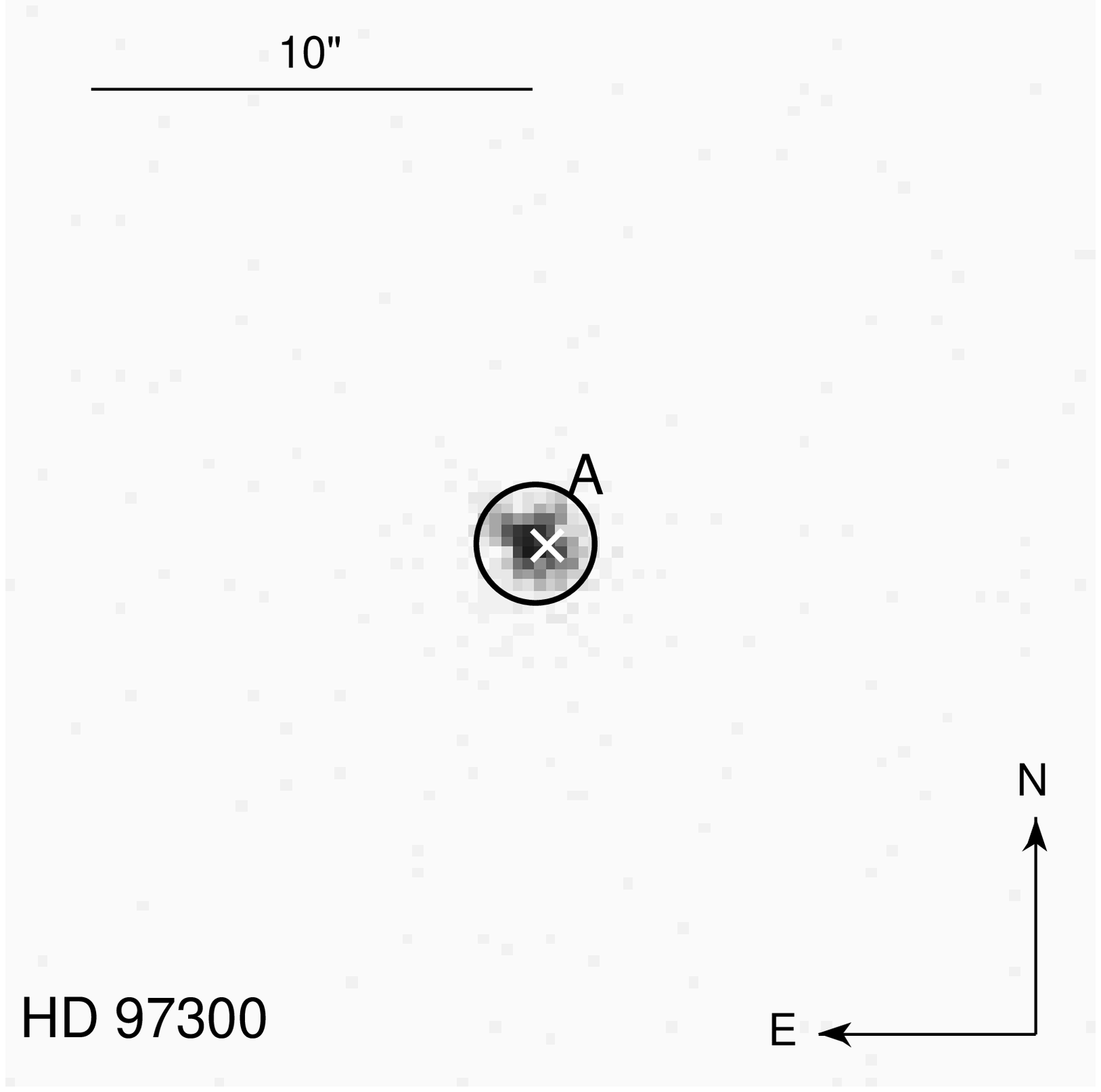}}
}
\parbox{6cm}{
\resizebox{6cm}{!}{\includegraphics{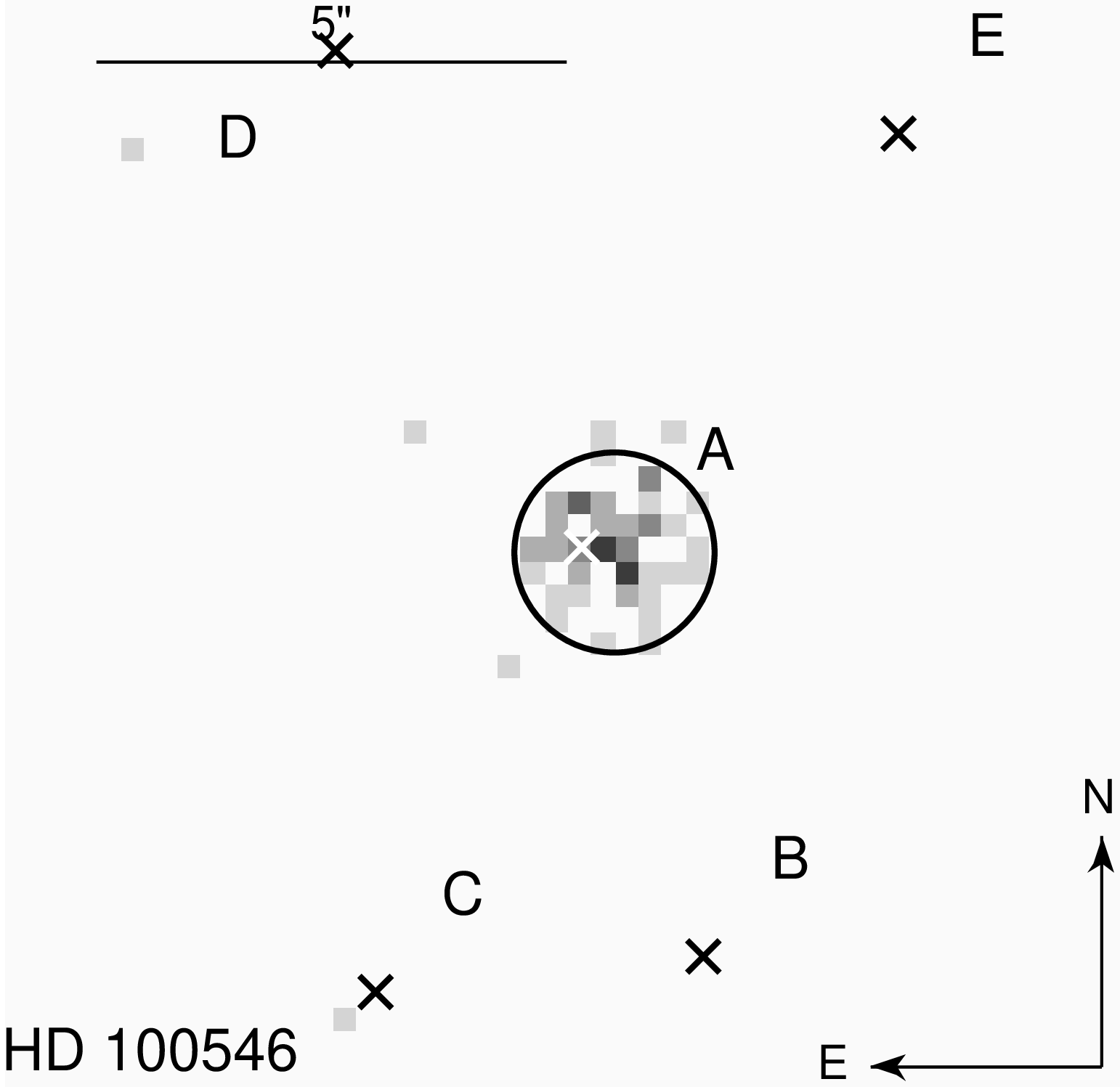}}
}
\parbox{6cm}{
\resizebox{6cm}{!}{\includegraphics{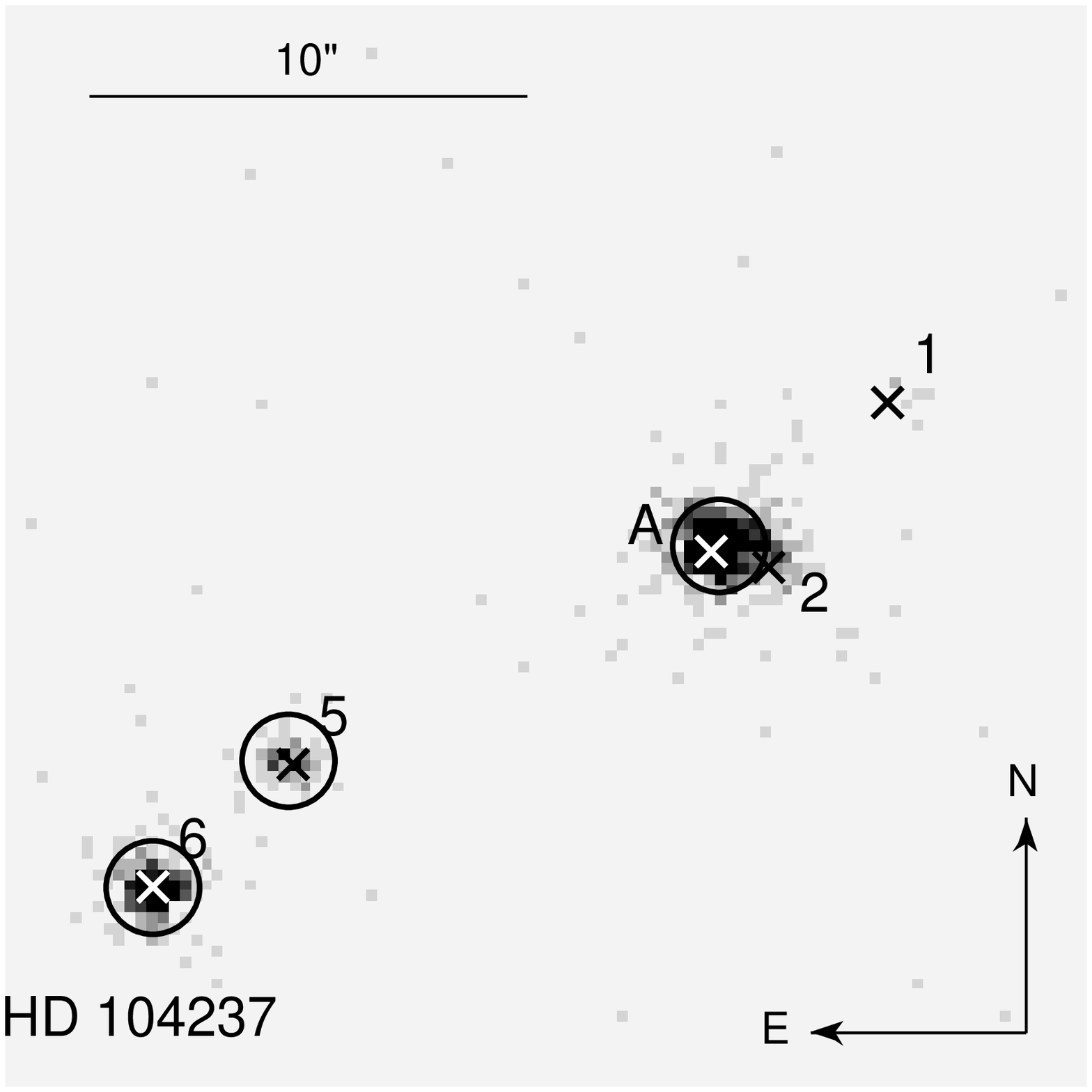}}
}
}
\parbox{18cm}{
\parbox{6cm}{
\resizebox{6cm}{!}{\includegraphics{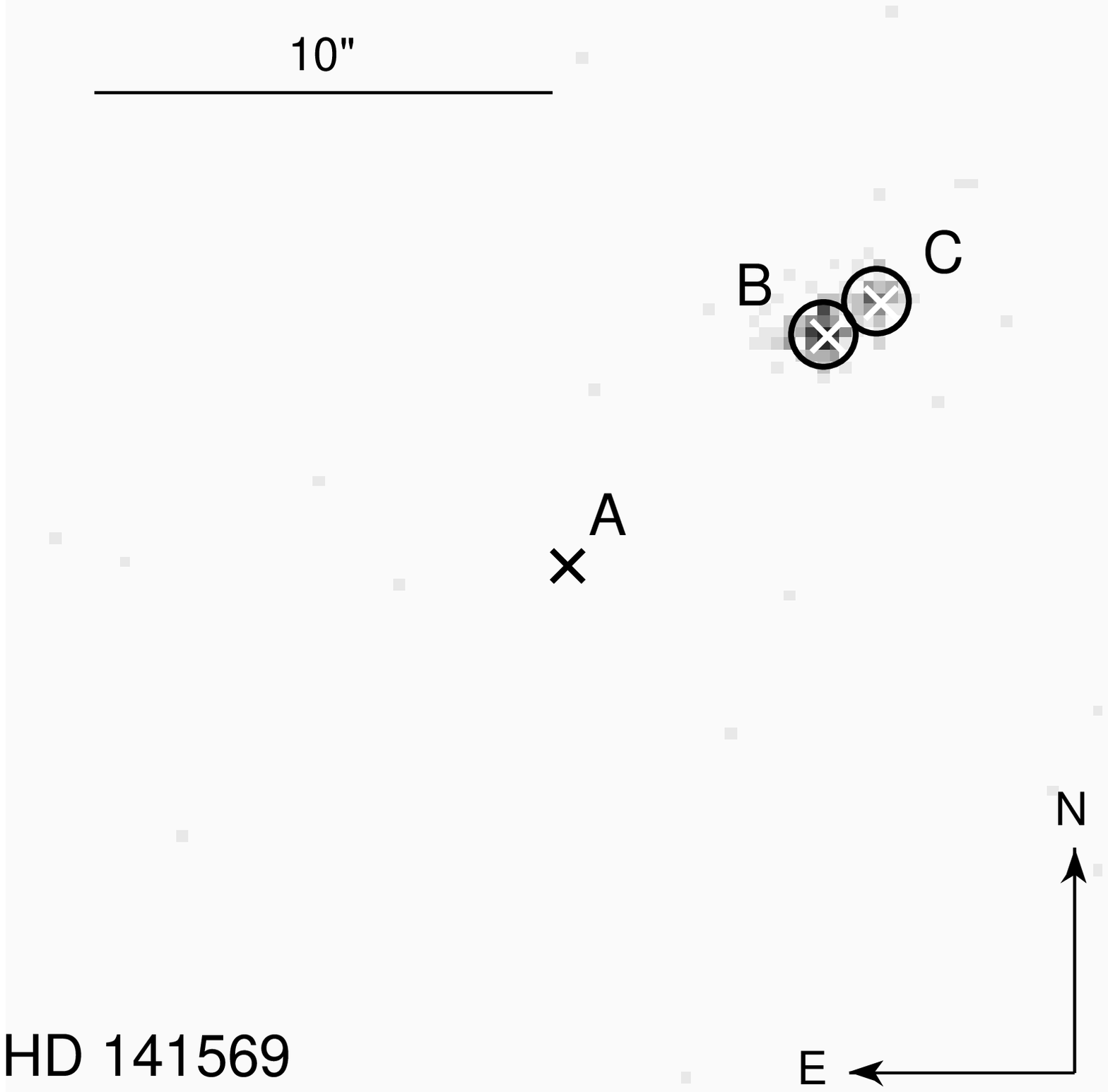}}
}
\parbox{6cm}{
\resizebox{6cm}{!}{\includegraphics{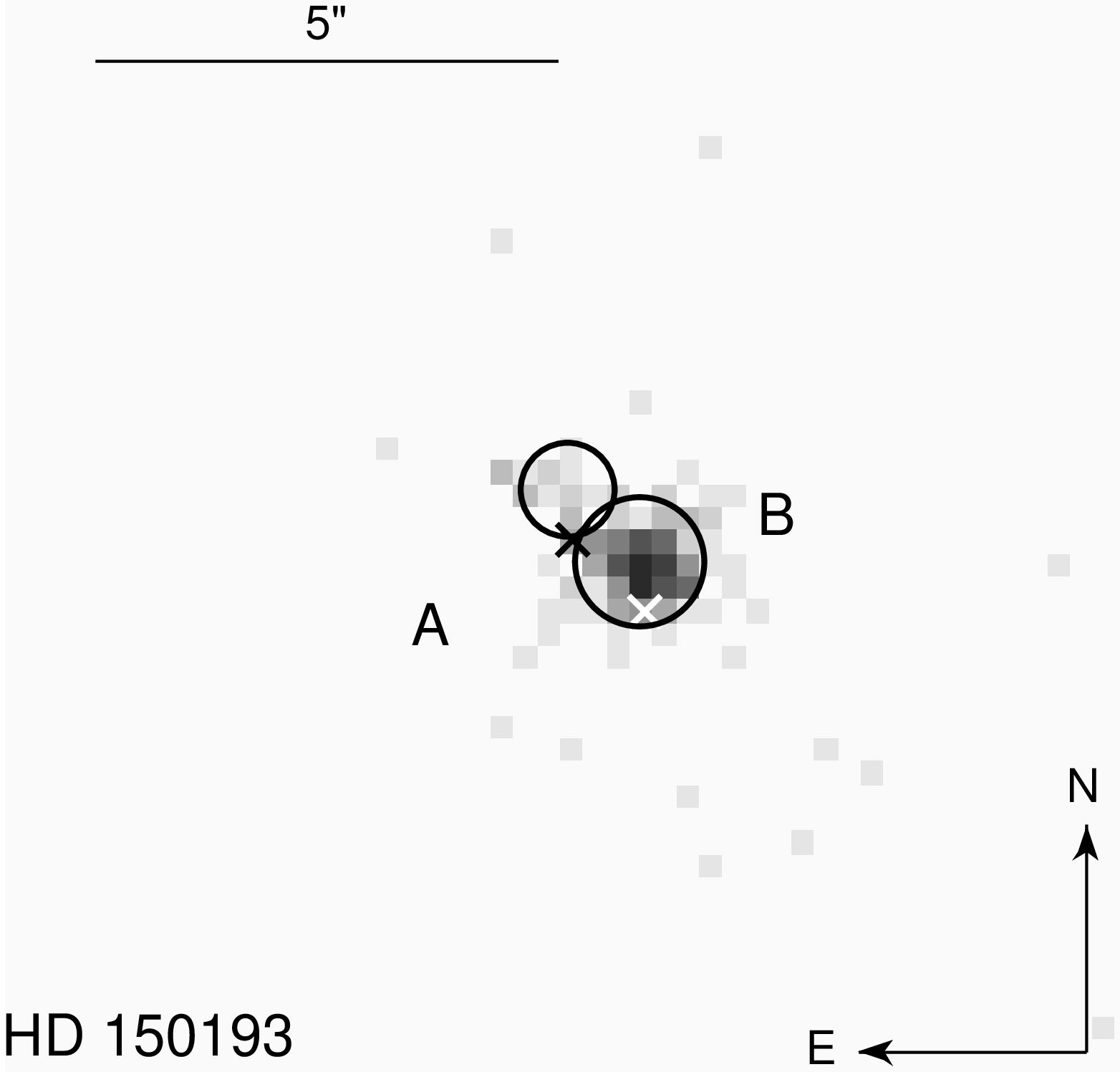}}
}
\parbox{6cm}{
\resizebox{6cm}{!}{\includegraphics{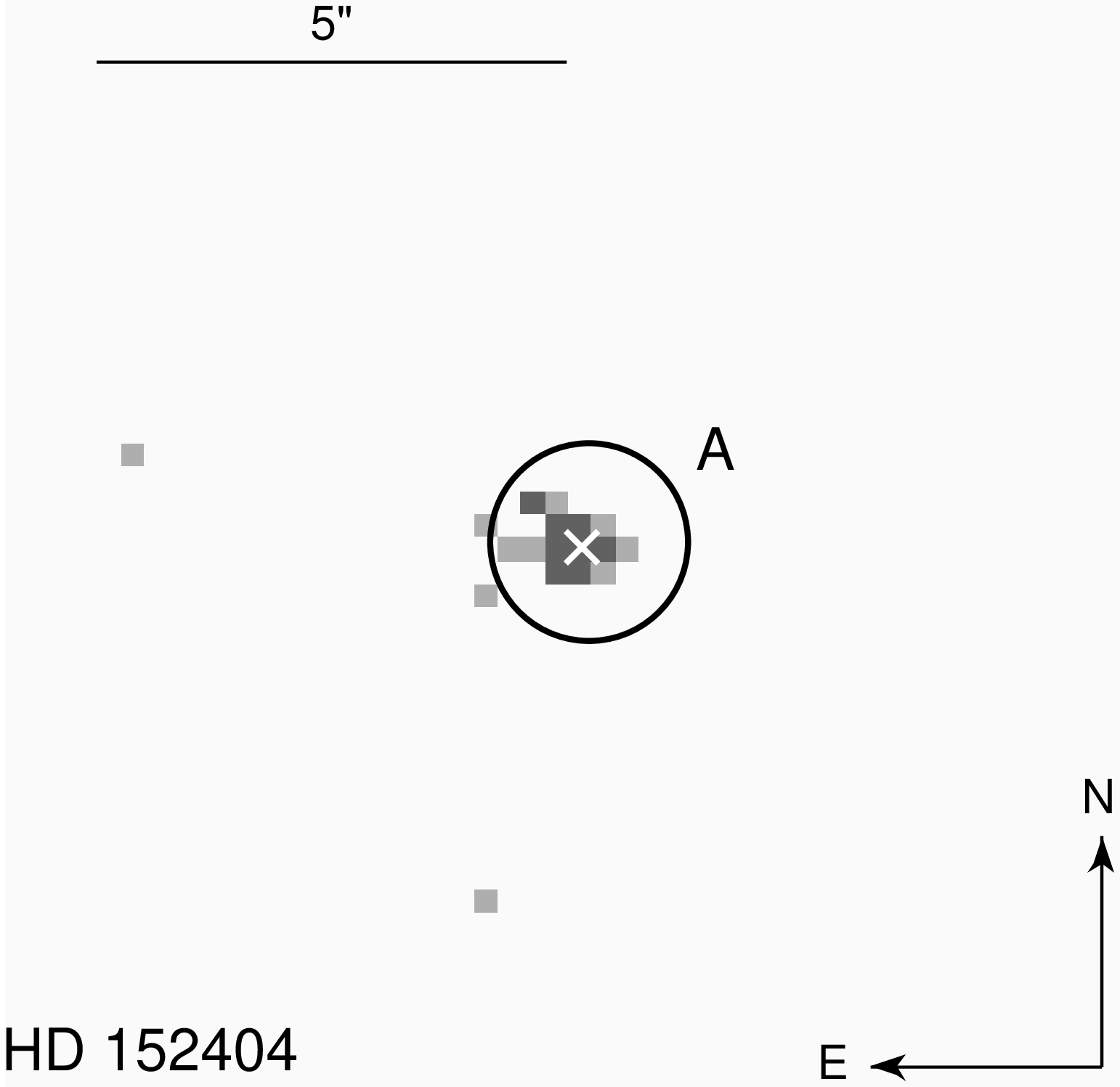}}
}
}
\caption{
{\em Chandra} ACIS images of HAeBe stars binned to a pixel size of $0.25^{\prime\prime}$.  
Crosses denote the optical/IR position of the individual components in the multiple system, 
circles mark the photon extraction areas centered on the position of X-ray sources detected
 with {\sc wavdetect}.}
\label{fig:acis_images_haebe}
\end{center}
\end{figure*}

\addtocounter{figure}{-1}

\begin{figure*}
\begin{center}
\parbox{18cm}{
\parbox{6cm}{
\resizebox{6cm}{!}{\includegraphics{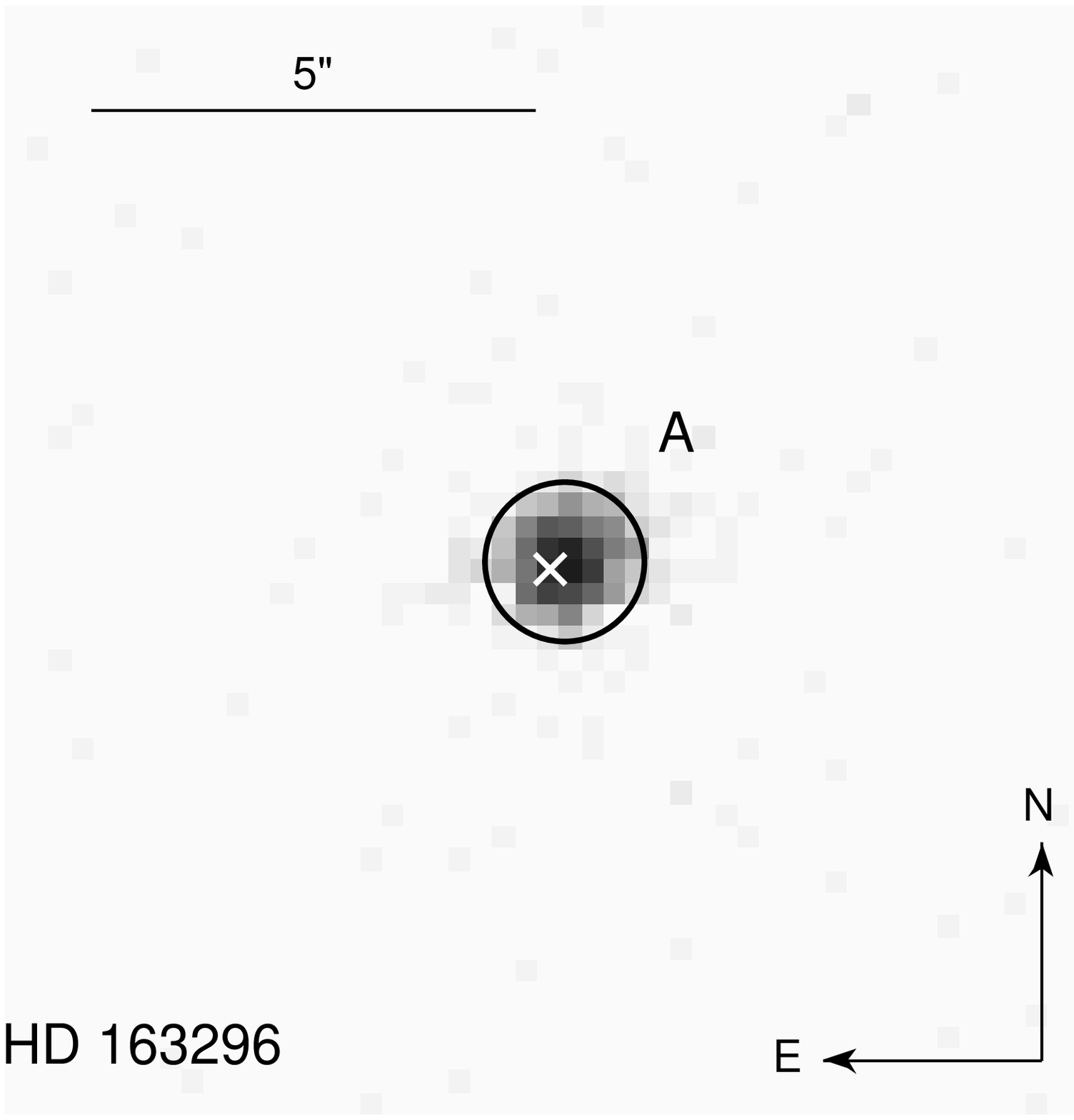}}
}
\parbox{6cm}{
\resizebox{6cm}{!}{\includegraphics{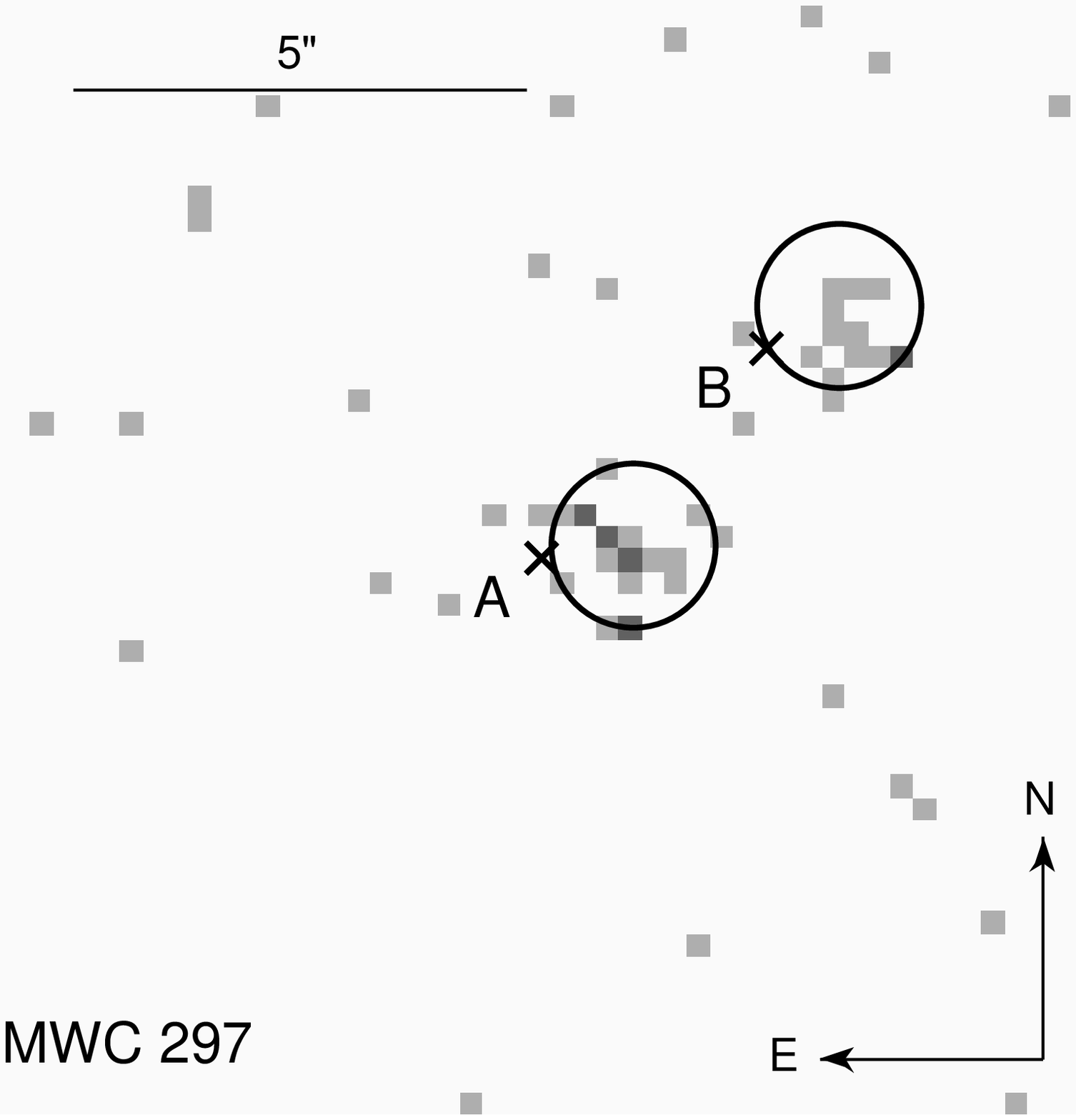}}
}
\parbox{6cm}{
\resizebox{6cm}{!}{\includegraphics{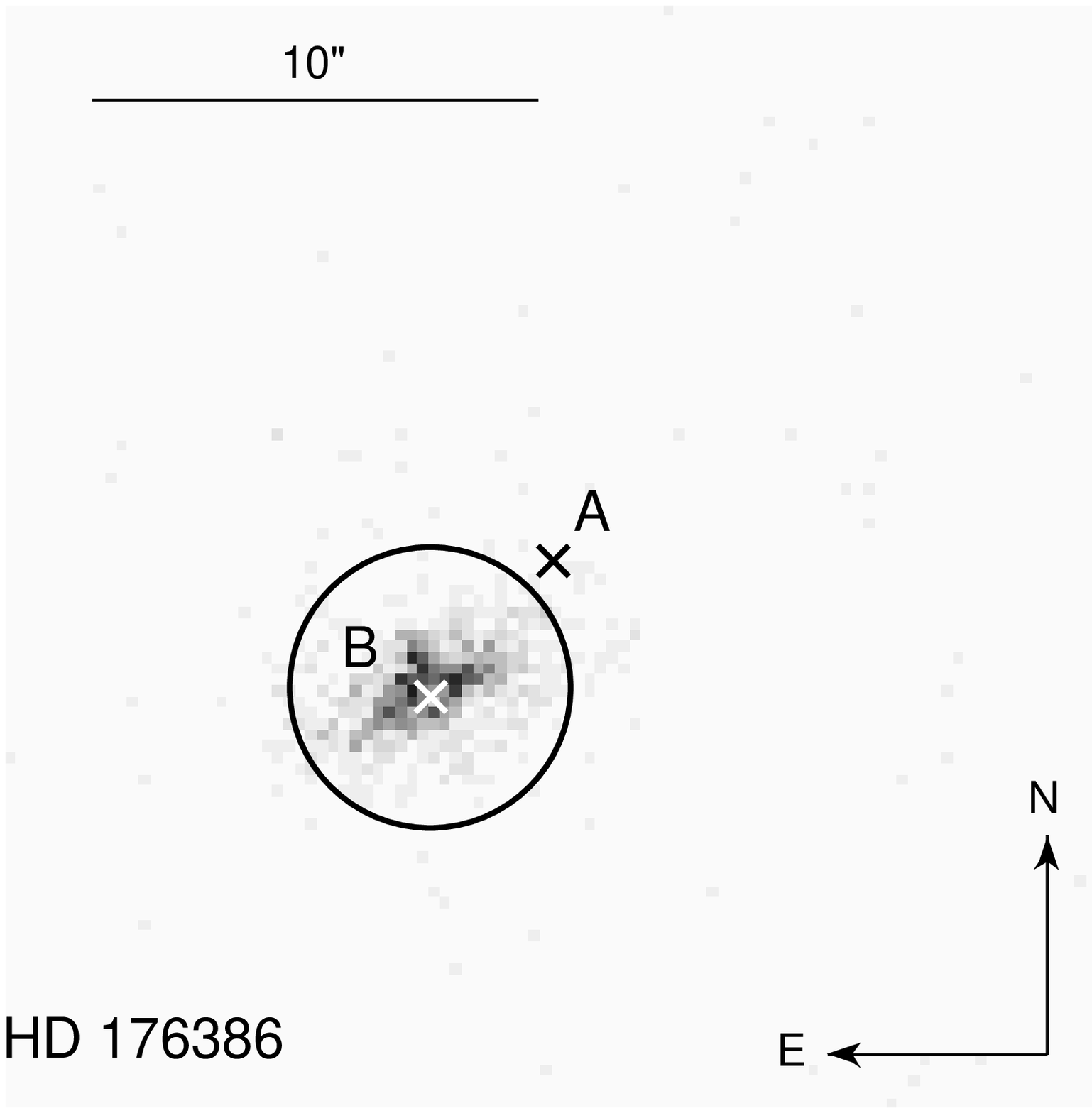}}
}
}
\parbox{18cm}{
\parbox{6cm}{
\resizebox{6cm}{!}{\includegraphics{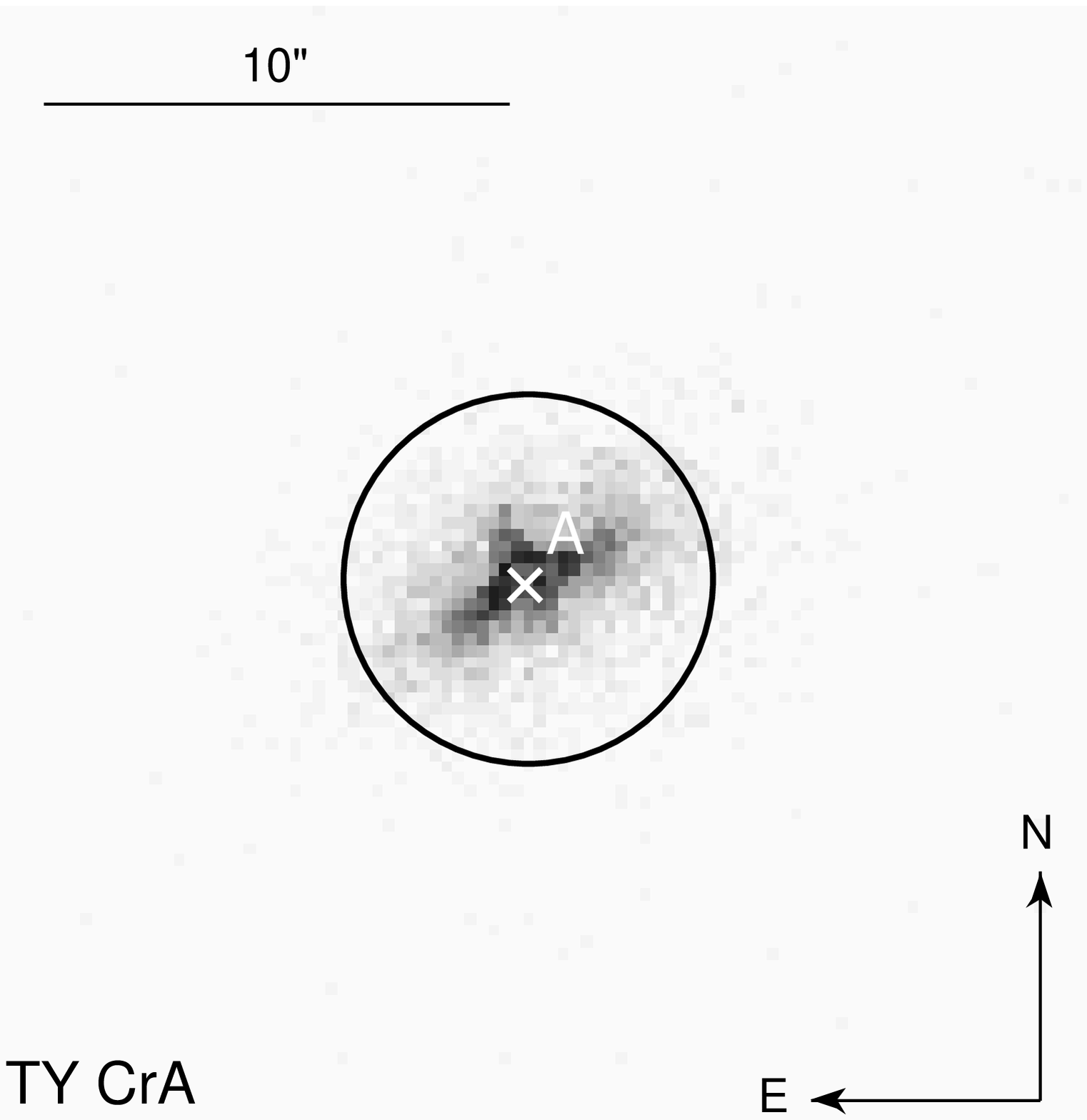}}
}
\parbox{6cm}{
\resizebox{6cm}{!}{\includegraphics{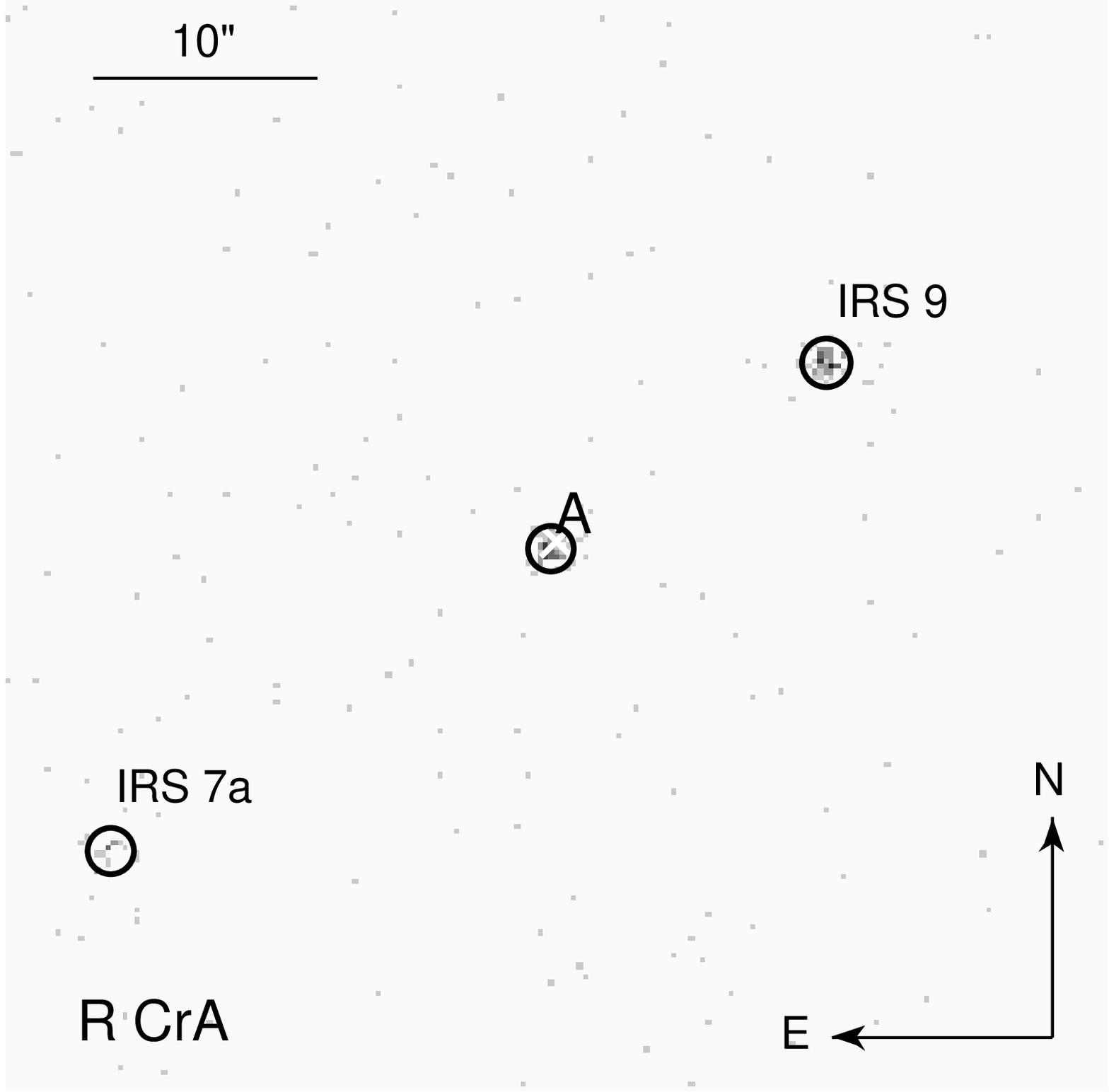}}
}
}
\caption{{\em continued}}
\end{center}
\end{figure*}

Source identification was done by measuring the separation of each 
X-ray source to the optical/IR positions of all known visual components. 
Then we assigned each X-ray source to the closest of the optical/IR objects. This 
procedure turned out to be unambiguous except in cases where the separation of
the components is below {\em Chandra}'s spatial resolving power, such that a single 
X-ray source represents more than one visual component. 
In the HD\,150193 system only one elongated X-ray source is detected, identified with
component~B, but the image clearly shows an enhancement of photons in the direction 
of component~A. Therefore, we consider HD\,150193\,A also as an X-ray source; see footnote of 
Table~\ref{tab:xrayparams_lx} for details. 
%
% OUTPUT FROM    write_xraytab.pro
%
\begin{sidewaystable*}\begin{center}
\caption{X-ray parameters of all components in the sample.}
\label{tab:xrayparams_lx}
\begin{tabular}{lccrrrrrrrrrrrr}\hline
Designation & Opt/IR & X-ray Iden. & Offax       & $\Delta_{\rm xo}$                       & Sign. & Counts$^{(1)}$ & \multicolumn{1}{c}{HR\,1} & \multicolumn{1}{c}{HR\,2} & PSF frac. & $\log{L_{\rm x}^{(1)}}$ & $\log{(L_{\rm x}^{(1)}/L_{\rm *})}$ & $P_{\rm KS}$ \\
            &        &             & [$^\prime$] & \multicolumn{1}{c}{[$^{\prime\prime}$]} &       &            &                           &                           & [\%]      & [erg/s]             &                                   & [\%]         \\ \hline
BD+30$^\circ$549 & A & X1& $   6.1 $ & $   0.36 $ & $   6.5 $ & $   43.0 \pm    7.6$ & $  0.63 \pm   0.23$ & $ -0.66 \pm   0.30$ & $ 0.90$ & $ 29.31$  & $ -5.21$  & $ 86.5$ \\
V892Tau  & A+C     & X1  & $   3.8 $ & $   0.99 $ & $  76.4 $ & $ 1180.0 \pm   35.4$ & $  0.92 \pm   0.02$ & $ -0.18 \pm   0.05$ & $ 0.90$ & $ 30.80$  & $ -2.65$  & $ 99.9$ \\
V892Tau  & B       & X2  & $   3.7 $ & $   0.68 $ & $  12.3 $ & $  170.0 \pm   14.1$ & $  0.81 \pm   0.07$ & $ -0.47 \pm   0.13$ & $ 0.90$ & $ 29.96$  & $ -3.11$  & $ 92.8$ \\
V380Ori  & A(SB)+B & X1  & $   2.0 $ & $   0.07 $ & $ 137.0 $ & $  888.0 \pm   30.8$ & $  0.65 \pm   0.04$ & $ -0.52 \pm   0.06$ & $ 0.90$ & $ 30.96$  & $ -4.67$  & $ 43.8$ \\
HD147889 & A(SB)   & $-$ & $   3.0 $ & $-$        & $-$       & $ <     4.4$         & $-$                 & $-$                 & $ 0.90$ & $< 28.55$ & $< -8.36$ & $ $ \\
V590Mon  & R\,2896 & X1  & $   2.6 $ & $   0.25 $ & $  57.2 $ & $  246.0 \pm   16.7$ & $  0.24 \pm   0.10$ & $ -0.68 \pm   0.13$ & $ 0.90$ & $ 30.51$  & $ -4.10$  & $ 90.2$ \\
V590Mon  & A       & $-$ & $   2.8 $ & $-$        & $-$       & $ <     4.4$         & $-$                 & $-$                 & $ 0.90$ & $< 28.76$ & $< -5.85$ & $ $ \\
ZCMa     &         & X1  & $   1.9 $ & $        $ & $   4.9 $ & $   12.0 \pm    4.6$ & $  0.67 \pm   0.56$ & $ -0.40 \pm   0.73$ & $ 0.91$ & $ 29.65$  & $      $  & $ 43.8$ \\
ZCMa     & A+B     & X2  & $   1.9 $ & $   0.16 $ & $  19.8 $ & $   53.0 \pm    8.3$ & $  0.43 \pm   0.22$ & $ -0.63 \pm   0.29$ & $ 0.91$ & $ 30.30$  & $ -8.43$  & $  6.7$ \\
HD97300  & A       & X1  & $   2.3 $ & $   0.27 $ & $ 130.8 $ & $ 1041.0 \pm   33.$  & $  0.58 \pm   0.04$ & $ -0.42 \pm   0.05$ & $ 0.90$ & $ 29.80$  & $ -5.33$  & $ 100.0$ \\
HD100546 & A       & X1  & $   1.7 $ & $   0.35 $ & $  21.3 $ & $   59.0 \pm    8.7$ & $  0.39 \pm   0.21$ & $ -0.61 \pm   0.28$ & $ 0.90$ & $ 28.93$  & $ -6.17$  & $  0.0 / 16.8^b $ \\
HD100546 & B       & $-$ & $   1.7 $ & $-$        & $-$       & $ <     4.4$         & $-$                 & $-$                 & $ 0.89$ & $< 27.81$ &           &   \\
HD100546 & C       & $-$ & $   1.6 $ & $-$        & $-$       & $ <     4.4$         & $-$                 & $-$                 & $ 0.89$ & $< 27.81$ &           &   \\
HD100546 & D       & $-$ & $   1.7 $ & $-$        & $-$       & $ <     4.4$         & $-$                 & $-$                 & $ 0.89$ & $< 27.81$ &           &   \\
HD100546 & E       & $-$ & $   1.6 $ & $-$        & $-$       & $ <     4.4$         & $-$                 & $-$                 & $ 0.89$ & $< 27.81$ &           &   \\
HD104237 & A(SB)   & X2  & $   1.7 $ & $   0.22 $ & $ 121.2 $ & $  757.0 \pm   28.5$ & $  0.27 \pm   0.05$ & $ -0.47 \pm   0.07$ & $ 0.90$ & $ 30.11$  & $ -4.90$  & $ 66.9 / 66.1^c $ \\
HD104237 & 1       & $-$ & $   1.7 $ & $-$        & $-$       & $ <    14.6$         & $-$                 & $-$                 & $ 0.89$ & $< 28.39$ &           &   \\
HD104237 & 2       & $-$ & $   1.7 $ & $-$        & $-$       & $ <    61.0$         & $-$                 & $-$                 & $ 0.58$ & $< 29.20$ &           &   \\
HD104237 & 5       & X3  & $   1.7 $ & $   0.13 $ & $  25.3 $ & $   69.0 \pm    9.4$ & $ -0.01 \pm   0.20$ & $ -0.88 \pm   0.28$ & $ 0.90$ & $ 29.06$  & $ -3.26$  & $ 83.8 / 18.7^c $ \\
HD104237 & 6       & X1  & $   1.7 $ & $   0.02 $ & $  62.7 $ & $  246.0 \pm   16.7$ & $  0.98 \pm   0.03$ & $  0.56 \pm   0.08$ & $ 0.90$ & $ 29.62$  & $ -2.92$  & $ 10.9 /  4.5^c $ \\
HD141569 & B       & X1  & $   1.6 $ & $   0.10 $ & $  51.3 $ & $  142.0 \pm   12.9$ & $  0.00 \pm   0.13$ & $ -0.69 \pm   0.19$ & $ 0.76$ & $ 29.64$  & $ -3.33$  & $ 33.9$ \\
HD141569 & A       & $-$ & $   1.7 $ & $-$        & $-$       & $ <     4.4$         & $-$                 & $-$                 & $ 0.89$ & $< 28.06$ & $< -6.80$ & $ $ \\
HD141569 & C       & X2  & $   1.6 $ & $   0.11 $ & $  24.5 $ & $   60.0 \pm    8.8$ & $ -0.13 \pm   0.22$ & $ -0.69 \pm   0.36$ & $ 0.76$ & $ 29.27$  & $ -3.37$  & $ 22.4$ \\
HD150193$^{(2)}$ & B       & X1  & $   1.7 $ & $   0.54 $ & $  55.4 $ & $  149.0 \pm   13.2$ & $  0.53 \pm   0.11$ & $ -0.49 \pm   0.15$ & $ 0.77$ & $ 30.22$  &           & $ 41.1$ \\
HD150193$^{(2)}$ & A(SB)   & X2  & $   1.7 $ & $   0.54 $ &           & $   14.0 \pm    4.8$ & $  0.29 \pm   0.55$ & $ -0.33 \pm   0.79$ & $ 0.58$ & $ 29.32$  & $ -5.65$  & $ 41.1$ \\
HD152404 & A(SB)   & X1  & $   1.7 $ & $   0.09 $ & $   8.3 $ & $   22.0 \pm    5.8$ & $ -0.27 \pm   0.41$ & $ -0.75 \pm   0.80$ & $ 0.90$ & $ 29.09$  & $ -5.45$  & $ 63.7$ \\
HD163296 & A       & X1  & $   0.3 $ & $   0.18 $ & $ 190.6 $ & $ 1056.0 \pm   33.5$ & $ -0.43 \pm   0.04$ & $ -0.87 \pm   0.08$ & $ 0.90$ & $ 29.60$  & $ -5.37$  & $ 99.8$ \\
MWC297   & A       & X1  & $   0.3 $ & $   1.02 $ & $   5.3 $ & $   16.0 \pm    5.1$ & $  0.75 \pm   0.42$ & $  0.00 \pm   0.58$ & $ 0.90$ & $ 29.28$  & $ -8.32$  & $ 36.9$ \\
MWC297   & B       & X2  & $   0.3 $ & $   0.93 $ & $   2.8 $ & $   11.0 \pm    4.4$ & $  1.00 \pm   0.44$ & $ -0.64 \pm   0.65$ & $ 0.90$ & $ 29.12$  &           & $ 39.7$ \\
HD176386 & B       & X1  & $   5.2 $ & $   0.21 $ & $  83.5 $ & $  954.0 \pm   31.9$ & $  0.23 \pm   0.05$ & $ -0.76 \pm   0.06$ & $ 0.90$ & $ 29.87$  &           & $ 98.5$ \\
HD176386 & A       & $-$ & $   5.2 $ & $-$        & $-$       & $ <    20.3$         & $-$                 & $-$                 & $ 0.39$ & $< 28.56$ & $< -6.72$ & $ $ \\
TYCrA    & A(SB)   & X1  & $   6.0 $ & $   0.15 $ & $ 134.8 $ & $ 4675.0 \pm   69.4$ & $  0.68 \pm   0.01$ & $ -0.40 \pm   0.02$ & $ 0.90$ & $ 30.63$  & $ -3.93$  & $100.0$ \\
RCrA     & IRS\,9  & X1  & $   1.7 $ & $   ?    $ & $  21.0 $ & $   53.0 \pm    8.3$ & $  1.00 \pm   0.08$ & $  0.09 \pm   0.24$ & $ 0.90$ & $ 28.76$  &           & $100.0$ \\
RCrA     & A       & X2  & $   1.7 $ & $   0.41 $ & $  20.7 $ & $   59.0 \pm    8.7$ & $  0.93 \pm   0.11$ & $  0.33 \pm   0.22$ & $ 0.90$ & $ 28.81$  & $ -4.59$  & $ 12.9$ \\
RCrA     & IRS\,7a & X3  & $   1.8 $ & $   ?    $ & $   5.1 $ & $   11.0 \pm    4.4$ & $  1.00 \pm   0.44$ & $  1.00 \pm   0.44$ & $ 0.90$ & $ 28.08$  &           & $ 81.4$ \\
\hline
\multicolumn{12}{l}{$^{(1)}$ in the $0.5-8$\,keV passband; $L_{\rm x}$ refers to the distance given in Table~\ref{tab:obslog} and has been corrected from the encircled PSF fraction.} \\
\multicolumn{12}{l}{$^{(2)}$ - Only one elongated source is detected with {\sc wavdetect}; photon extraction regions are defined as $0.7^{\prime\prime}$ radius circle on the position of the detected source} \\ 
\multicolumn{12}{l}{for component B, and as $0.5^{\prime\prime}$ radius circle centered on the second photon peak in the direction of the elongation for component A.} \\
\end{tabular}
\end{center}\end{sidewaystable*}

Table~\ref{tab:xrayparams_lx} summarizes the identification of all X-ray sources with components 
of our target systems and their X-ray parameters. We list also parameters for X-ray sources detected 
in the examined image portion but not known to be associated with the HAeBe targets.  
Cols.~1-5 give 
the designation of the target, % (column~$1$), 
component identifier, % (column~$2$),
X-ray source number, % (column~$3$), 
offaxis angle, 
and offset between X-ray and optical/IR position. % (column~$4$).  
For X-ray sources that are not 
known to be associated with the targets, we give an identifier in col.~2. 
Col.~6 represents the significance of the detection. 
The number of counts $N$ (column~$7$) refers to the $0.5-8$\,keV passband.
Errors were computed with the Gehrels approximation $\sqrt{N+0.75} + 1$ for
Poisson distributed data \citep{Gehrels86.1}. 
The irrelevant influence of the background is obvious from 
a look at the images in Fig.~\ref{fig:acis_images_haebe}. 
To quantify this statement, we estimated the background within a squared area of 
$1^\prime$ side length centered on the optical/IR position of the respective
star but excluding all detected sources. After scaling to the source extraction area this
background is negligibly small ($< 10^{-5}$\,cps), and therefore not considered in the analysis. 
As a rule, source photons were extracted from a circle centered on the
{\sc wavdetect} source position within the radius that includes 
$90$\,\% of the point-spread-function (PSF) at $1.5$\,keV, depending on the position 
of the source on the detector. For sources which are not fully resolved we defined smaller
non-overlapping circular photon extraction areas (see col.~10).
To compute upper limits for the undetected components of our target systems 
we used the method for Poisson-distributed counting data described by \citet{Kraft91.1}. 
Columns~$8$ and~$9$ of Table~\ref{tab:xrayparams_lx} show hardness ratios defined as 
$HR = (H-S)/(H+S)$, where $H$ and $S$ are the number of counts in a hard band
and in a soft band, respectively.
$HR1$ is defined using emission in the $0.5-1$\,keV ($S$) and the $1-8$\,keV ($H$) bands,
and $HR2$ from the $1-2$\,keV ($S$) and the $2-8$\,keV ($H$) band.
Col.~$10$ is the PSF fraction included in the source extraction area. 
In cols.~$11$ and~$12$ the PSF- and absorption-corrected X-ray luminosity
in the $0.5-8$\,keV band and the ratio of X-ray to bolometric luminosity are given. 
The X-ray luminosities have been computed
with PIMMS\footnote{The Portable Interactive Multi-Mission Simulator (PIMMS) is accessible at 
http://asc.harvard.edu/toolkit/pimms.jsp} 
assuming an iso-thermal emitting plasma with $kT=1$\,keV
and an absorbing column density $N_{\rm H}$ corresponding to the value derived from 
$A_{\rm V}$ according to the extinction law of \citet{Ryter96.1} 
($N_{\rm H}\,[10^{22}\,{\rm cm^{-2}}] = A_{\rm V}\,[{\rm mag}] \times 2.0 \times 10^{21}$). 
The bolometric luminosity represents the blackbody radiation from the stellar photosphere, 
without taking into account excess emission from circumstellar material seen at IR and radio
wavelengths; see Table~\ref{tab:early_hrd} for the 
stellar parameters of the HAeBe stars.
The assumption of an iso-thermal plasma may not be appropriate.  
However, we show in Sect.~\ref{subsect:spectra} that the X-ray luminosities derived in this way
are in reasonable agreement with the values obtained from the actual X-ray spectrum. 
Finally, col.$13$ of Table~\ref{tab:xrayparams_lx} represents the significance 
of variability according to the KS-test.

\subsection{X-ray lightcurves}\label{subsect:lcs}

\begin{figure*}
\begin{center}
\parbox{18cm}{
\parbox{9cm}{
\resizebox{9cm}{!}{\includegraphics{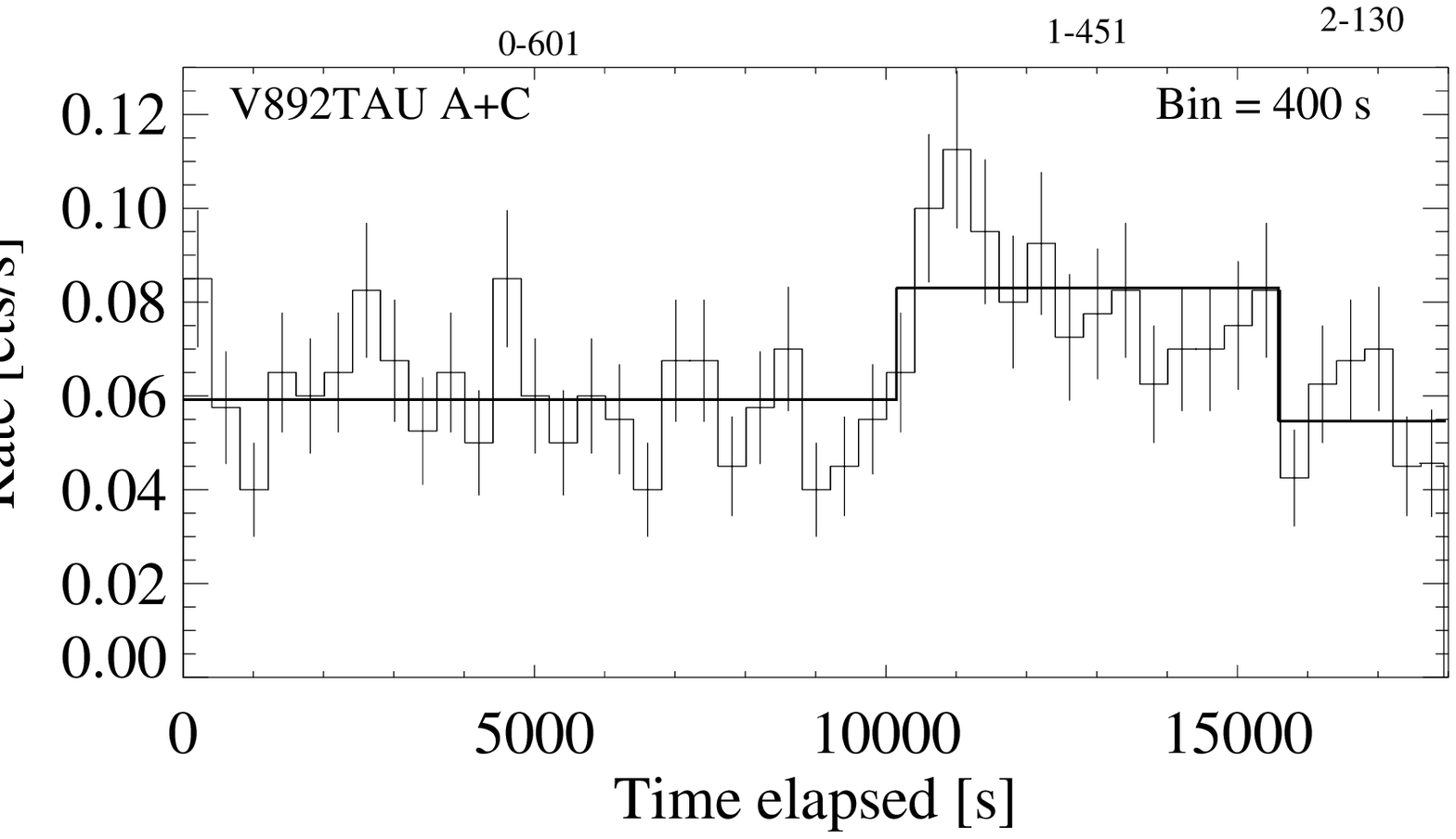}}
}
\parbox{9cm}{
\resizebox{9cm}{!}{\includegraphics{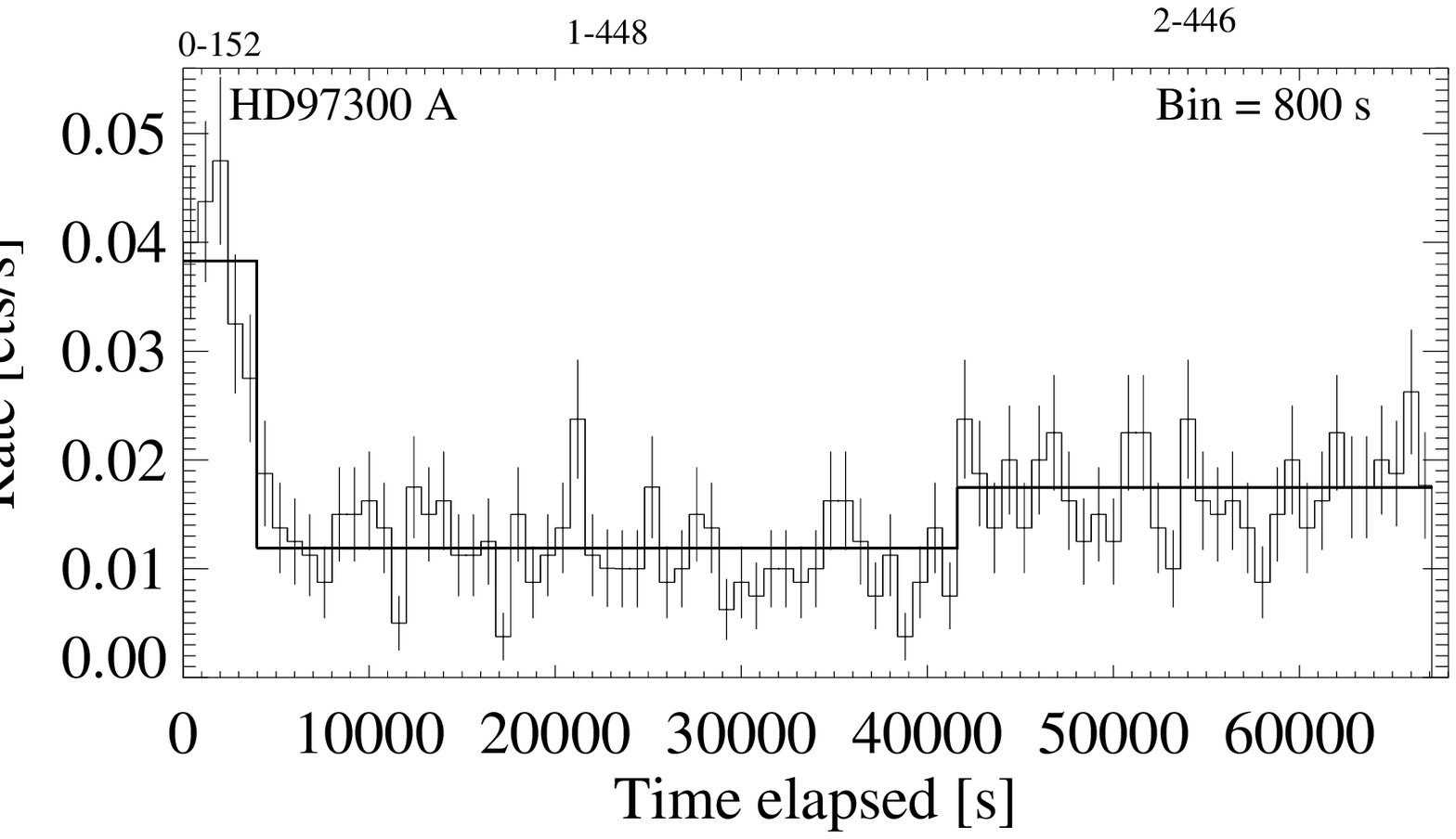}}
}
}
\parbox{18cm}{
\parbox{9cm}{
\resizebox{9cm}{!}{\includegraphics{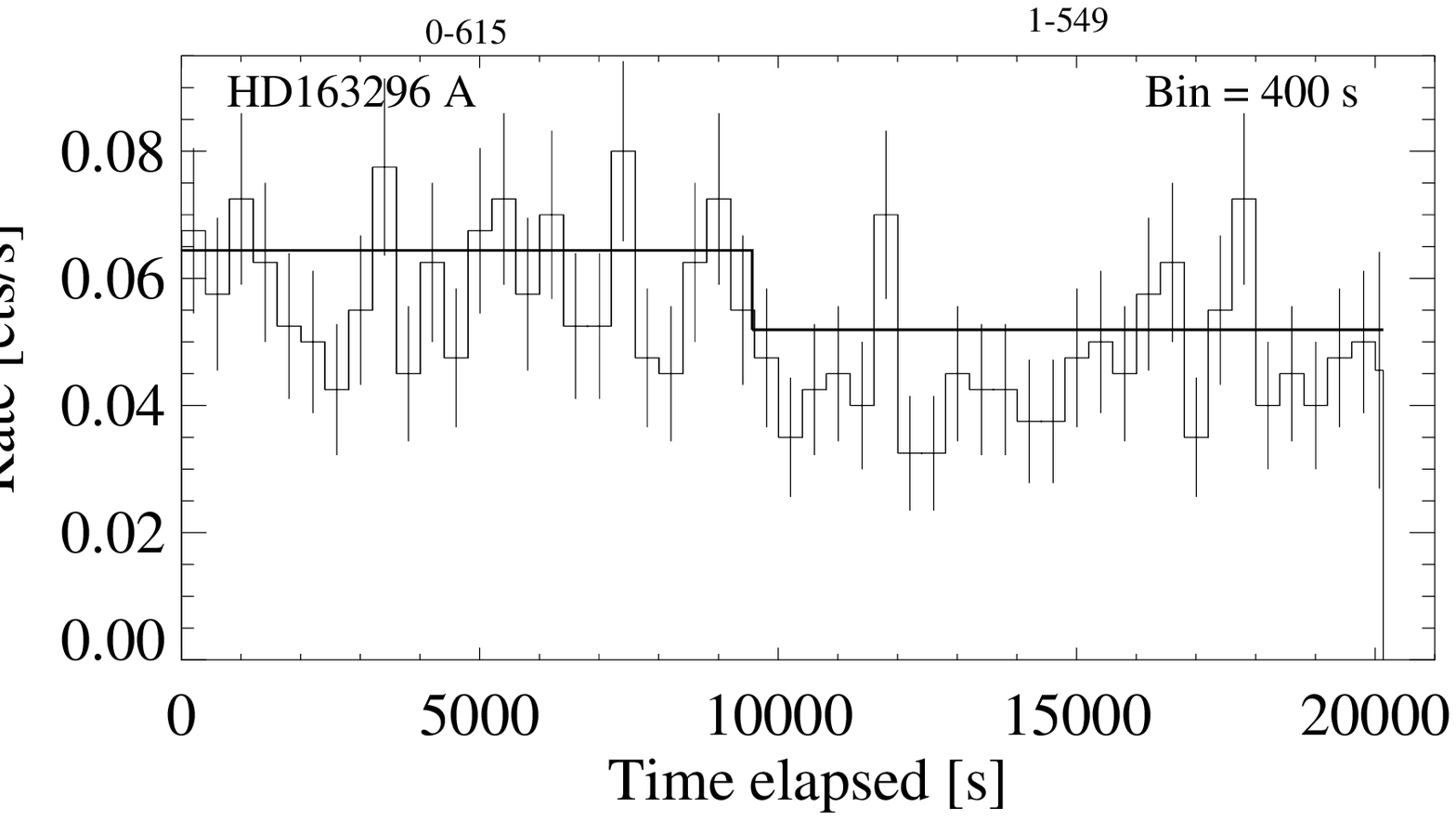}}
}
\parbox{9cm}{
\resizebox{9cm}{!}{\includegraphics{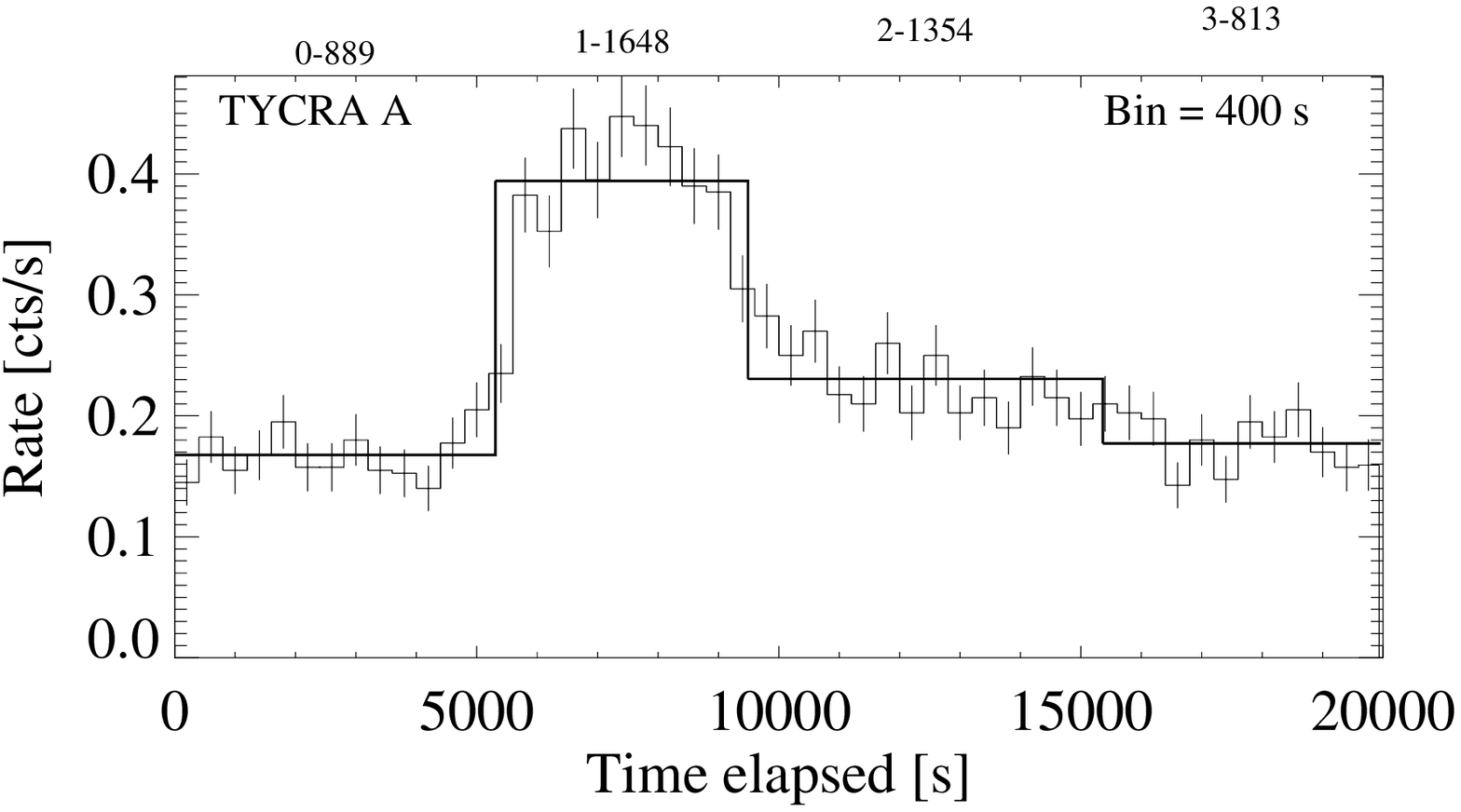}}
}
}
\parbox{18cm}{
\parbox{9cm}{
\resizebox{9cm}{!}{\includegraphics{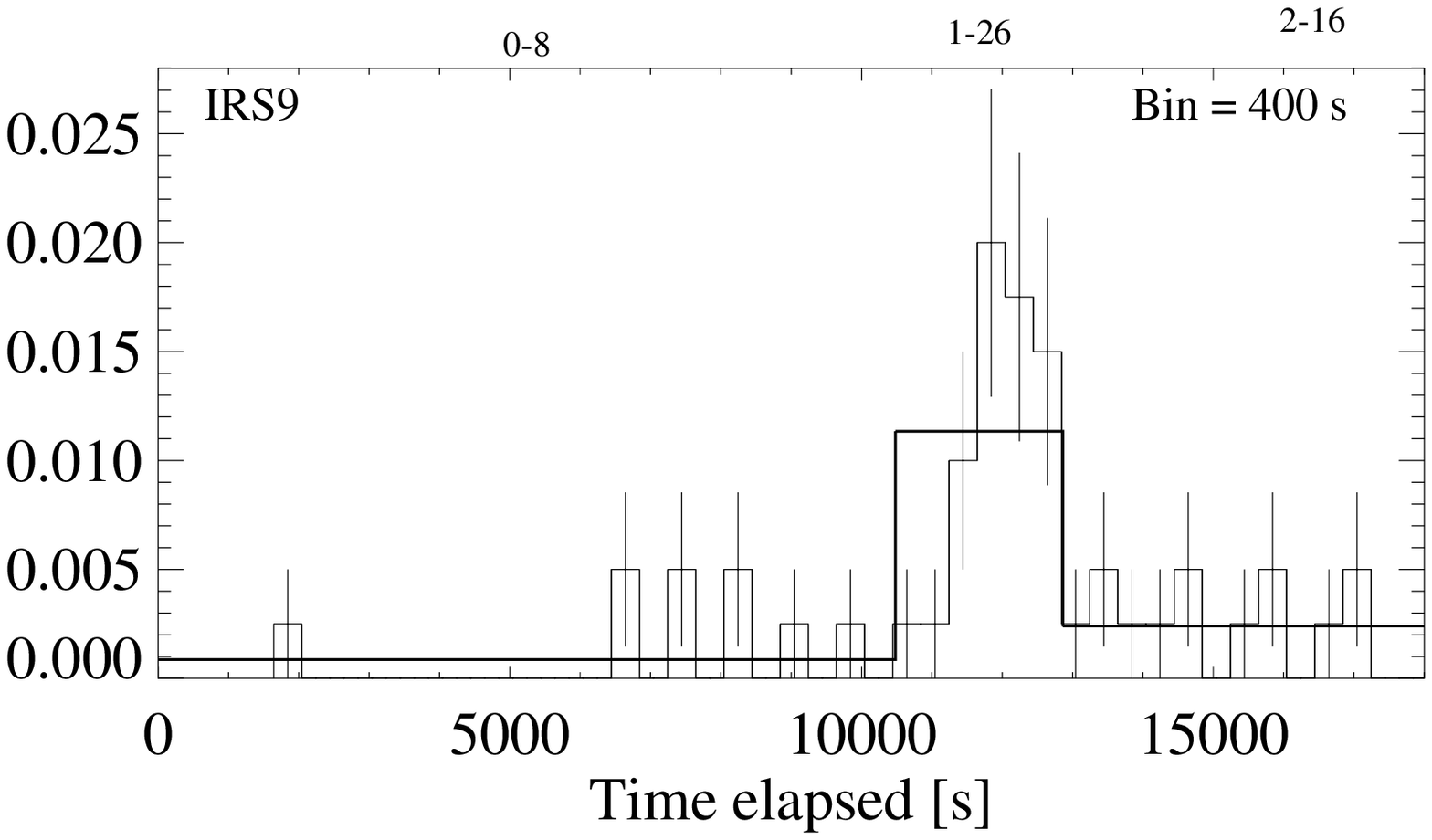}}
}
\parbox{9cm}{
\caption{{\em Chandra} ACIS lightcurves of sources for which a KS-test yields variability at the $99$\,\% 
confidence level. Overlaid on the binned lightcurves are the blocks resulting from the MLB analysis;
see text in Sect.~\ref{subsect:lcs}. A sequential number for the blocks and the number of counts in each 
block is given on top of the figure.}
\label{fig:acis_lcs}
}
}
\end{center}
\end{figure*}

Lightcurves were binned for all detected sources into $400$\,s and $800$\,s intervals.
Obviously, the detection of variability requires both high statistics and a good time
resolution, and our choice of the binning is a compromise between these two
opposing effects. The probabilities for variability according to the KS-test 
are summarized in the last column of Table~\ref{tab:xrayparams_lx}. 
HD\,100546 and HD\,104237 were observed twice, and the time-series of both data sets
were analysed separately, to avoid mixing long and short timescales.
The detection of variability (at the $99$\,\% confidence level) in only $5$ of the $29$ examined
time series may be due to the short exposure times of some observations. 
In Fig.~\ref{fig:acis_lcs} the lightcurves of the variable sources are displayed.  

As a further test on variability we applied a %maximum likelihood `blocking' (MLB) 
method that determines periods of constant signal from a list of photon arrival times, 
based on a maximum likelihood algorithm under the assumption of Poisson statistics. 
This technique is termed MLB (maximum likelihood blocking) henceforth, 
see \citet{Wolk05.1} for a a more detailed description.  
The algorithm has two free parameters: the minimum
number of counts accepted in a given segment and the significance,
defined as the $1 - P $, where $P$ is the probability that a constant lightcurve is
spuriously segmented due to random fluctuations. We set the significance to $99$\,\%,
and the minimum number of counts in a segment was $20$, except for weak sources,
where we allowed as few as a single count to define a segment. The segmented lightcurves
of the sources found to be variable with the KS-test are overplotted on the binned 
curves in Fig.~\ref{fig:acis_lcs}. 

We find excellent agreement between the detection
of variability with the KS-test and with the MLB method. 
All stars variable according to the KS-test are also found to be variable with the MLB technique,
while all stars non-variable according to the KS-test are described by a single segment in
the MLB analysis. 
With the exception of TY\,CrA the luminosity is not significantly affected by flaring,
and we do not evaluate flare and quiescent data separately.

\subsection{X-ray spectra}\label{subsect:spectra}

% -- TABLE WITH RESULTS FROM SPECTRAL FITS
%
%    COMBINATION FROM xrayspectab_1tapec
%                     xrayspectab_1tapec_freeze_nh
%                     xrayspectab_2tapec
%                     xrayspectab_2tapec_freeze_nh
%
\begin{table*}\begin{center}
\caption{Spectral parameters of X-ray sources associated with HAeBe stars and their known close companions. The ``='' sign indicates fits with column densities fixed on the value derived from the optical extinction; see text in Sect.~\ref{subsect:spectra}.}
\label{tab:xrayspectab}
\begin{tabular}{lccrrrrrrrr}\hline
HD & Opt/IR & X-ray Iden. & $\chi^2_{\rm red}$ (dof) & $P_{\rm Null}$ & $N_{\rm H}$                & $kT_1$ & $kT_2$ & $\log{EM_1}$       & $\log{EM_2}$       & $\log{L_{\rm x}}^*$ \\
   &        &             &                          &                & [$10^{22}\,{\rm cm^{-2}}$] & [keV]  & [keV]  & [${\rm cm}^{-3}$] & [${\rm cm}^{-3}$] & [erg/s]          \\ 
\hline
\multicolumn{11}{c}{HAeBe stars} \\
\hline
BD$+30^\circ549$ & A       & X-1 &   $0.83$ ($  5$) &    $0.526$ & $  1.87$ &   $0.3$ & $-$    &  $53.9$ & $-$    &  $30.4$ \\
V892\,Tau        & A+C     & X-1 &   $0.99$ ($ 47$) &    $0.493$ & $  1.36$ &   $2.0$ & $-$    &  $53.7$ & $-$    &  $30.7$ \\
V380\,Ori        & A(SB)+B & X-1 &   $1.62$ ($ 43$) &    $0.006$ & $  0.48$ &   $1.6$ & $-$    &  $54.4$ & $-$    &  $31.3$ \\
Z\,CMa           & A+B     & X-2 &   $0.29$ ($  7$) &    $0.957$ & $  0.29$ &   $1.2$ & $-$    &  $53.6$ & $-$    &  $30.5$ \\
HD\,97300        & A       & X-1 &   $1.13$ ($ 39$) &    $0.269$ & $= 0.27$ &   $0.8$ & $ 2.5$ &  $51.9$ & $52.8$ &  $29.9$ \\
HD\,104237$^a$   & A(SB)   & X-2 &   $1.34$ ($ 21$) &    $0.138$ & $= 0.06$ &   $0.8$ & $ 4.0$ &  $53.1$ & $53.0$ &  $30.3$ \\
HD\,104237$^b$   & A(SB)   & X-1 &   $1.31$ ($ 17$) &    $0.172$ & $= 0.06$ &   $0.7$ & $ 5.1$ &  $53.1$ & $52.7$ &  $30.2$ \\
HD\,163296       & A       & X-1 &   $1.53$ ($ 33$) &    $0.027$ & $= 0.05$ &   $0.5$ & $10.9$ &  $52.9$ & $51.9$ &  $29.7$ \\
TY\,CrA          & A(SB)   & X-1 &   $1.03$ ($128$) &    $0.383$ & $  0.49$ &   $0.8$ & $ 3.2$ &  $53.5$ & $53.7$ &  $30.9$ \\
R\,CrA           & A       & X-2 &   $0.72$ ($  8$) &    $0.675$ & $  1.11$ &   $>2.8$ & $-$   &  $51.9$ & $-$    &  $29.1$ \\
\hline
\multicolumn{11}{c}{Companions and unrelated objects} \\
\hline
V892\,Tau        & B       & X-2 &   $0.66$ ($ 13$) &    $0.801$ & $  0.76$ &   $1.7$ & $-$ &  $52.8$ & $-$ &  $29.7$ \\
HD\,104237$^a$   & 6       & X-1 &   $1.97$ ($ 11$) &    $0.027$ & $  3.58$ &   $2.2$ & $-$ &  $53.6$ & $-$ &  $30.7$ \\
HD\,141569       & B       & X-1 &   $0.91$ ($  9$) &    $0.513$ & $  0.07$ &   $0.8$ & $-$ &  $52.9$ & $-$ &  $29.7$ \\
HD\,141569       & C       & X-2 &   $0.47$ ($  8$) &    $0.879$ & $  0.00$ &   $0.8$ & $-$ &  $52.4$ & $-$ &  $29.2$ \\
HD\,150193       & B       & X-1 &   $1.19$ ($ 11$) &    $0.288$ & $  0.15$ &   $1.5$ & $-$ &  $53.3$ & $-$ &  $30.1$ \\
HD\,176386       & B       & X-1 &   $1.09$ ($ 45$) &    $0.318$ & $= 0.12$ &   $1.0$ & $-$ &  $53.1$ & $-$ &  $29.9$ \\
R\,CrA           & IRS\,9  & X-1 &   $0.92$ ($  7$) &    $0.489$ & $  1.36$ &   $1.5$ & $-$ &  $52.0$ & $-$ &  $29.2$ \\
\hline
\multicolumn{11}{l}{$^*$ in the $0.5-8$\,keV passband; $L_{\rm x}$ is absorption corrected and refers to the distance given in Table~\ref{tab:obslog}} \\
\multicolumn{11}{l}{$^a$ in Obs-ID\,3428, $^b$ in Obs-ID\,2404} \\
\end{tabular}
\end{center}\end{table*}

The preparations for the spectral analysis consist of the following steps: 
we extracted a spectrum, the corresponding detector response matrix that maps
pulse heights into energy space, and an
auxiliary response file which contains information about the effective area
and detector efficiency across the chip as a function of energy.
Then we binned each spectrum to a minimum of $5$ or more counts per bin starting at
$0.5$\,keV.  
As the background of ACIS is very low ($< 1$\,count in the source extraction area)
it can be neglected. 
For several X-ray sources in the sample presented here the number of counts collected 
is rather small, and the results are discussed only for the $17$ sources
with at least five degrees of freedom after binning the spectrum and choosing an adequate model. 
Since the models have at least $2$ free parameters (see below), all analysed sources have more than
$35$ counts. In practice, the lowest S/N spectrum (BD$+30^\circ549$) has $>40$ counts. 
Spectral modelling was performed in the XSPEC environment, version 11.3.0. 

We tried four different models on each spectrum: one- and two-temperature thermal models
with fixed or freely variable absorption from interstellar and circumstellar material. 
The thermal components are represented by 
{\sc APEC}\footnote{For a description of the Astrophysical Plasma Emission Code ({\sc APEC}) see \cite{Smith01.1}.} 
with atomic cross-sections and elemental abundances from \citet{Wilms00.1}.
For the fits with fixed absorption, the column density $N_{\rm H}$ of the {\sc WABS} component 
was set to the values derived from the visual extinction assuming the \citet{Ryter96.1} extinction law, 
and the column density of the companions is assumed to be equal to that of the HAeBe stars, 
i.e. all components in a given system are considered to be hidden behind the same amount of 
interstellar plus circumstellar absorbing material.
Global metal abundances were fixed to $0.3$ times the solar value. 

The minimum number of free parameters is two ($kT$, $EM$),  
and the maximum is five ($N_{\rm H}$, $kT_1$, $kT_2$, $EM_1$, $EM_2$). 
For each model spectral fitting was started on several choices for the initital parameters, 
and then the fit with the minimum $\chi^2$ was adopted. Then we decided for each
star individually which of the four models represents best the data. 
We base the evaluation of the quality of a fit result on the null probability that the spectrum is described
by the model. Generally, simpler models are considered first, 
because an excessive number of free parameters with respect to the data statistics does not 
yield physical constraints \citep[cf. discussion by ][]{Flaccomio06.1}. 
If $P_{\rm Null} > 5$\,\% we accept the 1-T model with free column density  
(1T)\footnote{A few exceptions do not satisfy our criterion of $P_{\rm Null} > 5$\,\% in
any of the four models. 
HD\,104237-6 is among the fainter sources and shows erratic jumps in the spectrum.
For the rather bright X-ray sources corresponding to V380\,Ori and HD\,163296 
the discrepancies between the observed spectrum and the models can probably
be attributed to peculiar abundance patterns. Since we are interested in the global properties of HAeBe stars
rather than a detailed examination of individual sources, we judge these fits as acceptable after 
visual inspection.}. 
Otherwise, we move on to the 1-T model with fixed $N_{\rm H}$ (1T$_{\rm Av}$). 
The preference of 1T versus 1T$_{\rm Av}$ despite its smaller degrees of freedom is empirical:  
The 1T model is the one that gives on the sample average better fits than the 1T$_{\rm Av}$ model,
probably because 1T$_{\rm Av}$ sometimes ends up in a local minimum of the $\chi^2$-space.  
An example for such a case is BD$+30^\circ549$, whose best-fit model has too large $N_{\rm H}$
compared with its $A_{\rm V}$, and, to compensate for this, has untypically small $kT$ and large $L_{\rm x}$ 
(see also Sect.~\ref{subsect:spec_pec}). 
The preference of 1T is also motivated by the fact that 
there are considerable uncertainties in the visual extinction of the targets,
such that even under the assumption of the \citet{Ryter96.1} extinction law 
the column density is not well known. 
If none of the 1-T models is satisfactory, the 2-T model with fixed $N_{\rm H}$ is chosen (2T$_{\rm Av}$). 
Finally, the last option is the 2-T model with free absorption (2T). 

The parameters derived from spectral fitting are presented in 
Table~\ref{tab:xrayspectab}. Luminosities refer to the $0.5-8.0$\,keV band after
correction for absorption. 
A comparison of the X-ray luminosities derived from
integrating the X-ray spectrum and those calculated from the count rate with PIMMS
shows that the values lie within a factor of two of each other, with one exception
(BD$+30^\circ549$) discussed below.  
This proves that our simple estimate with PIMMS yields 
reliable X-ray luminosities even if the assumed spectral parameters may not reflect precisely
the conditions in the emitting plasma. 
Throughout the remainder of this paper both luminosities will be used:
in the direct comparison of parameters from the X-ray spectrum $L_{\rm x,SPEC}$ is used. 
Otherwise $L_{\rm x,PIMMS}$ is preferred because it enables to consider all targets,  
including faint sources and non-detections.

\subsection{Peculiar spectral properties}\label{subsect:spec_pec}

The majority of stars are characterized by X-ray temperatures of $\sim 1-2.5$\,keV, 
similar to earlier results from ASCA \citep{Hamaguchi05.1},  
and moderate extinction. 
A deviation from this general picture is seen in four sources: the HAeBe stars BD$+30^\circ549$ and R\,CrA, 
the Class\,I protostar IRS\,9 in the R\,CrA star forming complex, 
and the probably accreting T\,Tauri star HD\,104237-6,  
are strongly absorbed ($N_{\rm H} > 10^{22}\,{\rm cm^{-2}}$).  
Except for BD$+30^\circ549$, these stars show also rather high temperature. 
All of them are relatively faint, such that the shape of their spectrum cannot 
be examined in more detail. 

For BD$+30^\circ549$ the typical degeneracy of $N_{\rm H}$ with $kT$ often observed in spectra with
low statistics, may have led to a wrong estimate
of both parameters (cf. discussion in the previous paragraph). 
Nevertheless, we retain this result as best fit because fixing the absorbing 
column yielded a much poorer $\chi^2$ and residuals. 
  
IRS\,9 has flared (see Fig.~\ref{fig:acis_lcs}), and HD\,104237-6 has shown a drastic change in count rate 
by a factor of $8$ between the two {\em Chandra} exposures. The spectrum of this star refers to the
exposure Obs-ID\,3428, 
where HD\,104237-6 was probably caught in a flare-like high state. These flares may partly explain
the high observed temperatures. Another possibility to keep in mind is, that in the presence of 
strong absorption the soft spectral component may become unrecognizable.  
The temperature of R\,CrA is unconstrained, and we give the lower limit in Table~\ref{tab:xrayspectab}.

\section{Discussion}\label{sect:discussion}

\subsection{The nature of the X-ray emitters}\label{subsect:xray_nature}

The HAeBe stars observed with {\em Chandra} do not represent an unbiased sample, because
(i) some of them are picked up based on X-ray detections in previous X-ray surveys, and 
(ii) they cover a vast range of distances and do not form a volume limited sample.
X-ray emission for a distance limited, small sample of HAeBe stars has been 
studied by \citet{Skinner04.1}, but this group is not complete either and did not comprise 
non-detections.  
Furthermore, mixing of data obtained with different instruments may introduce uncertainties
in the X-ray parameters. 
Finally, we call the controversial definition of the HAeBe class to the reader's attention.
Our sample was selected from published catalogs that contain also some dubious cases, e.g. two 
stars of our sample have spectral type F 
(Z\,CMa and HD\,152404), and -- strictly speaking -- are not Herbig Ae/Be objects. For more details
on the nature of the individual stars we refer to the appendix. 
Having these caveats in mind, we can proceed to some 
considerations on the {\em Chandra} observations of HAeBe stars. 

A total of $13$ out of $17$ HAeBe stars are detected. Seven of them have 
(visual and/or spectroscopic) companions unresolvable with {\em Chandra} imaging, 
that might be responsible for the X-ray emission.   
Five HAeBe stars observed with {\em Chandra} have no earlier X-ray data. 
The detection fraction of the HAeBe stars in this sample  
is $\sim 76$\,\% if the known or hypothesized 
companions that remain unresolvable are not taken into consideration, $\sim 53$\,\% if
the emission is attributed to known unresolved {\it visual} companions,
and $\sim 35$\,\% if the emission is attributed to known unresolved {\it visual or spectroscopic}
companions (cf. Table~\ref{tab:early_hrd}). 
This result suggests that either a large fraction of HAeBe stars %(and MS B-stars) 
are intrinsic X-ray sources, or a substantial fraction of companions has yet to be discovered. 

Similar conclusions on the detection fractions 
were reached for MS B-type stars observed with {\em Chandra} \citep{Stelzer06.1}. 
However, a direct comparison with the detection fractions of HAeBe stars is impeded by 
the different criteria for the sample selection. 
The MS B star sample is composed of previously known X-ray emitters for which the presence of 
faint companions within $1-8^{\prime\prime}$ had been demonstrated before, while for the HAeBe sample
no systematic survey for companions exists.  

\begin{figure}
\begin{center}
\resizebox{9cm}{!}{\includegraphics{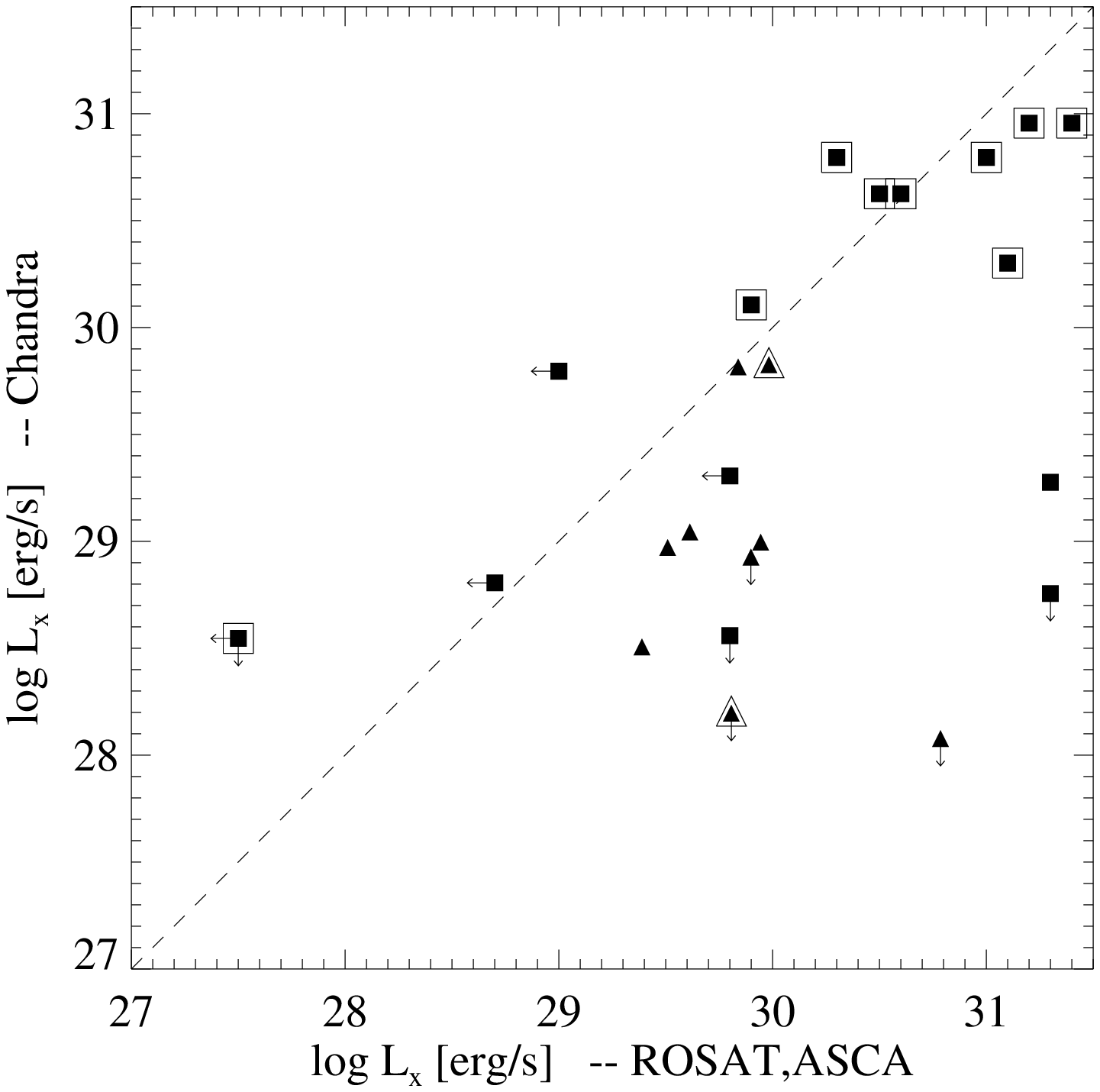}}
\caption{X-ray luminosities for HAeBe stars (squares) and MS B-type stars (triangles) 
measured with different instruments. 
{\em x-axis} - $L_{\rm x}$ from {\em ROSAT} or {\em ASCA}, 
{\em y-axis} - $L_{\rm x}$ from {\em Chandra}. 
All luminosities refer to the $0.5-8$\,keV band. 
{\em ROSAT} luminosities are from \protect\citet{Zinnecker94.1} and \protect\citet{Hamaguchi05.1} 
for HAeBe stars and from \protect\citet{Berghoefer96.1} for MS B-type stars; 
{\em ASCA} luminosities for HAeBe stars are from \protect\citet{Hamaguchi05.1};
{\em Chandra} luminosities for MS B-type stars are from \citet{Stelzer06.1} 
Binaries that are unresolved with {\em Chandra} are surrounded by larger symbols. 
Three stars (the topmost in terms of {\em Chandra} luminosity) appear twice in this diagram because 
they have both {\em ROSAT} and {\em ASCA} data.}
\label{fig:lx_new_old}
\end{center}
\end{figure}

Interestingly, despite the unprecedented $1^{\prime\prime}$ resolution of {\em Chandra},  
we can establish for only two HAeBe stars (V590\,Mon and HD\,176386) 
that previous lower spatial resolution observations erroneously considered them X-ray emitters
due to confusion with other objects. 
(It is assumed here that the non-detections are X-ray dark. This seems justified by the rather 
stringent upper limits we derive on the X-ray luminosity.) 
For many of the stars observed with {\em ROSAT} and/or {\em ASCA} 
the low spatial resolution data have yielded surprisingly good estimates for the 
X-ray luminosity, although in other cases $L_{\rm x}$ was overestimated due to source confusion. 
This is shown in Fig.~\ref{fig:lx_new_old} which compares the X-ray luminosities derived from 
{\em Chandra} with those from {\em ROSAT} and {\em ASCA}.
Our {\em Chandra} imaging has verified that objects located below the diagonal in 
the diagram were severely contaminated by nearby sources in the {\em ROSAT} and {\em ASCA} observations. 
%
% OUTPUT FROM    plot_specparams.pro
%

Strong constraints are placed on the X-ray luminosity of all four undetected HAeBe stars:
$\log{L_{\rm x}} < 28.8$\,erg/s. This is consistent with the absence of unknown companions
to these stars, because virtually all TTS with age of $< 10$\,Myr emit at higher levels;
see e.g. \citet{Stelzer00.1}.
In line with this argument, all but one of the four non-detections are not known to be multiple. 
The exception (HD\,147889) is a spectroscopic binary, but composed of two B2-type stars.

\subsection{X-ray Properties}\label{subsect:xray_prop}

To examine the X-ray characteristics of the HAeBe stars and their companions,
they are compared to other samples of intermediate- and low-mass stars
observed with {\em Chandra}: the above mentioned MS B-type stars, and young
stars in the Orion Nebula Cluster (ONC). An extensive X-ray study of the ONC was performed 
in the framework of the {\em Chandra} Orion Ultradeep Project (COUP); 
see ApJ Supplement Special Issue 160.  

\begin{figure}
\begin{center}
\resizebox{9cm}{!}{\includegraphics{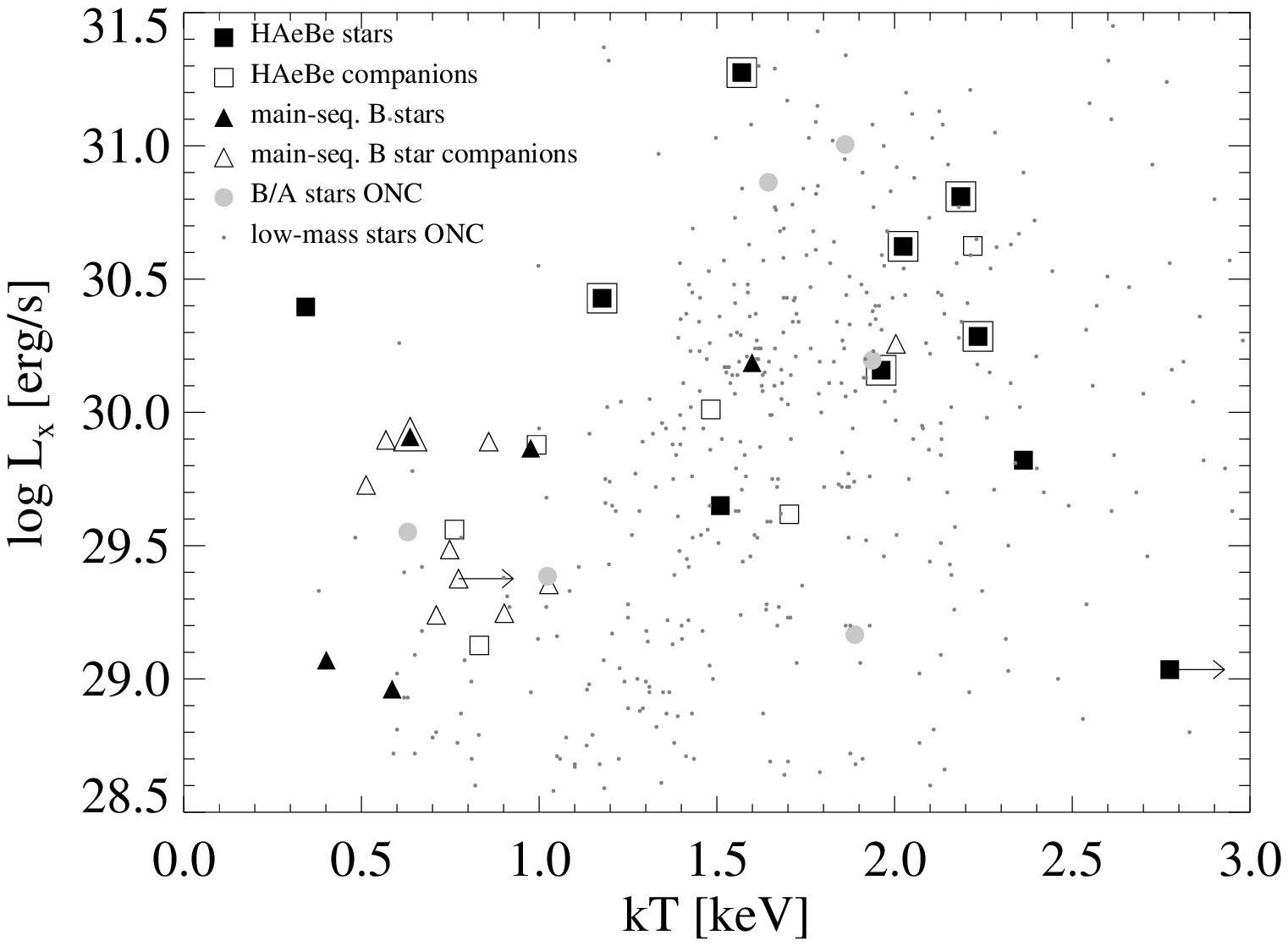}}
\caption{X-ray temperature and luminosity for HAeBe stars and their companions 
compared to main-sequence B stars and companions,  
and to young stars in Orion; see text Sect.~\ref{subsect:xray_prop}. 
Binaries that are unresolved with {\em Chandra} are surrounded by larger symbols.}
\label{fig:tx_lx}
\end{center}
\end{figure}

For the stars with 2-T spectra we computed a mean coronal temperature, defined as the average
of the two temperatures weighted with the emission measures. 
Fig.~\ref{fig:tx_lx} displays the X-ray luminosity versus X-ray temperature
for the HAeBe stars (filled squares), the MS B-type stars (filled triangles) 
and the ONC B/A-type stars (filled circles).
Lower mass companions to each of these groups are represented by open plotting symbols. 

The distribution of the HAeBe systems in the $\log{L_{\rm x}} - kT$ plot more
closely resembles that of young, low-mass stars in the ONC than that of MS B star systems. 
While the B stars on the MS, as well as most of their companions, are characterized
by $kT \sim 0.5...1$\,keV, the HAeBe stars show a much larger range with the majority of them
at $kT > 1$\,keV. 
For the only exception of a HAeBe star with a low temperature, 
BD$+30^\circ549$ (already discussed in Sect.~\ref{subsect:spec_pec}), 
the fit is ambiguous. Recall, that the 1T model was chosen as best fit based on statistics. 
The 1T$_{\rm Av}$ model would place this star at a more typical temperature of $kT \sim 1.3$\,keV, and  
lower the luminosity to $\log{L_{\rm x}} \sim 29.2$\,erg/s. % in Fig.~\ref{fig:tx_lx}. 

The {\em Chandra} observations presented here allow us for the first time a systematic 
direct comparison between the X-ray properties of HAeBe stars and their presumably coeval companions.
There is some inconclusive indication, that the resolved HAeBe companions are softer and 
less luminous than their primaries. On the other hand, if the unresolved HAeBe companions are
also considered X-ray emitters, the distribution is again similar to stars in the low-mass ONC sample.  
%
% OUTPUT FROM    plot_specparams.pro
%

A comparison of the variability of X-ray sources associated with HAeBe stars and those ascribed to
their resolved companions may, in principle, be employed to discriminate between intrinsic 
X-rays from HAeBe stars and the companion hypothesis. In the sample of six variable sources, 
two are apparently single, two are HAeBe stars with spectroscopic or
unresolved visual companion, one is a resolved companion, and one is an unrelated object (IRS\,9). 
The different lengths of the observations and the different intensities of the
sources lead to a vast variety of lightcurves. 
As far as we can tell, there is no clear distinction between the primaries and the secondaries,
but this assertion needs to be corroborated by more extensive X-ray data.

\subsection{Fractional X-ray Luminosity}\label{subsect:xray_lbol}

\begin{figure}
\begin{center}
\resizebox{9cm}{!}{\includegraphics{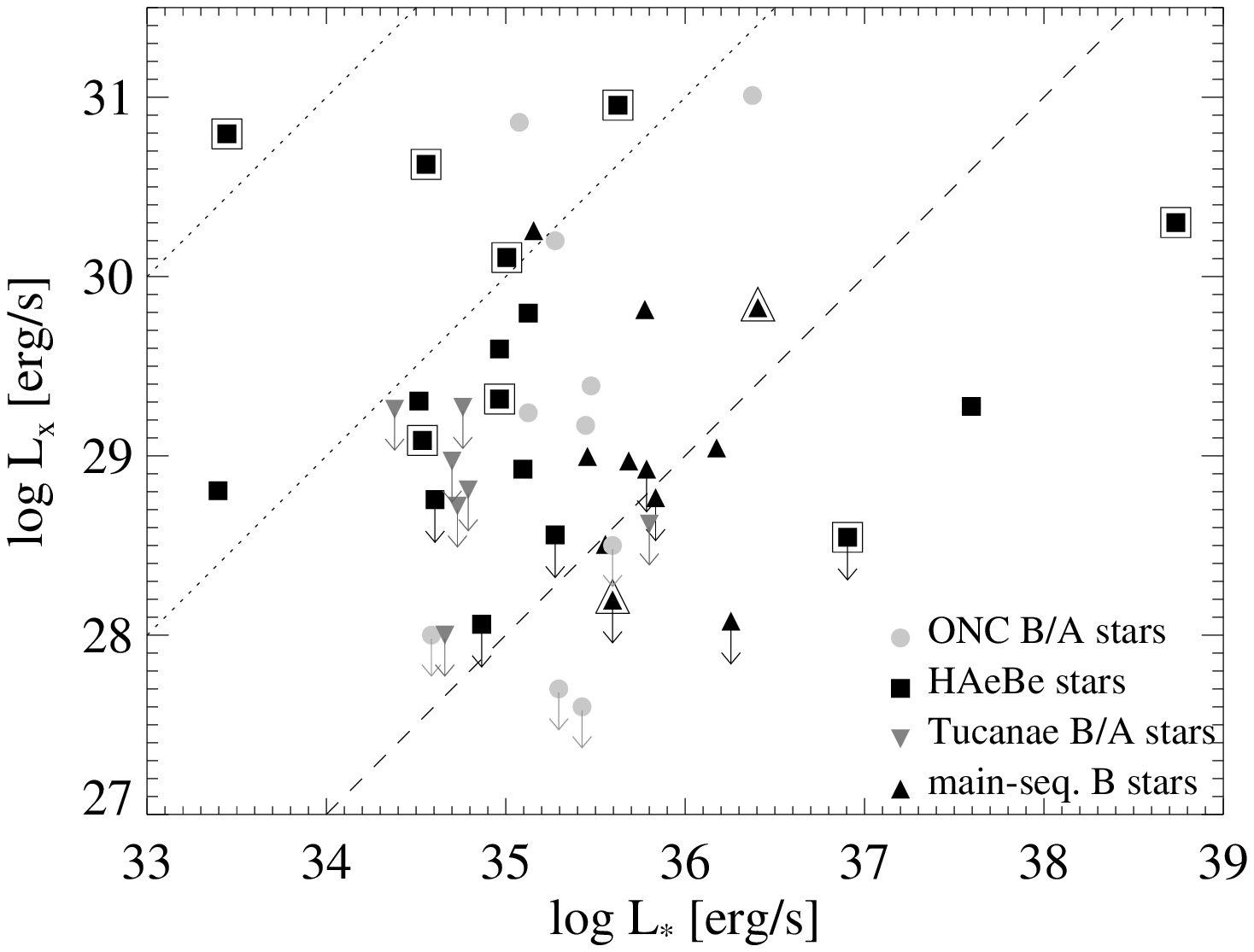}}
\caption{X-ray versus stellar luminosity for HAeBe stars and other samples of intermediate-mass stars. 
The X-ray data derived from {\em Chandra} with the exception of Tucanae, where we use {\em ROSAT} All-Sky
Survey observations. The dashed line indicates a value of $L_{\rm x}/L_{\rm bol} = 10^{-7}$,
the dotted lines represent $10^{-5}$ and $10^{-3}$, respectively.} 
\label{fig:lx_lbol}
\end{center}
\end{figure}

The $L_{\rm x}/L_{\rm *}$ ratio is examined in Fig.~\ref{fig:lx_lbol} 
%for the HAeBe stars and the MS B stars. 
for several samples of intermediate-mass stars in different evolutionary phases. 
As some HAeBe stars are still surrounded by 
luminous circumstellar matter, it is important to distinguish between bolometric and
stellar luminosity. For stars on the MS $L_{\rm *}$ and $L_{\rm bol}$ are identical. 

The HAeBe stars are compared to the $\sim 1$\,Myr old BA stars in the ONC introduced
above \citep[see also ][]{Stelzer05.1}, 
the $\sim 30$\,Myr old BA stars in the Tucanae association \citep{Stelzer00.1}, and the 
MS B stars also introduced above. The latter ones represent on average a slightly 
hotter and optically brighter sample than the others, because they do not include A-type stars.

A few HAeBe stars have been reported to
be very luminous, with $L_{\rm *} > 10^{37}$\,erg/s, similar to the range for O stars.
However, we note that the stellar luminosities of some HAeBe stars are highly uncertain
(cf. Table~\ref{tab:early_hrd}). 
In particular, the situation for Z\,CMa is confusing, with vastly different measurements
for $L_{\rm *}$ given in the literature. Throughout this paper we have used  
$\log{(L_{*}/L_\odot)} = 5.15$ following \citet{Acke04.1};
see Appendix~\ref{subsect:zcma} for more details on the stellar parameters of Z\,CMa. 
In any case, 
this object is peculiar in our HAeBe sample, because it is highly embedded. 

The bulk of the HAeBe stars show $L_{\rm *} \sim 10^{34...35}$\,erg/s, and
are characterized by a $3$\,dex
spread in X-ray luminosity, with values significantly lower than predicted by the
$L_{\rm x}/L_{\rm *}$ relation of TTS. 
This was already obvious from the restricted sample presented
by \citet{Skinner04.1}. 
Two stars, R\,CrA and V892\,Tau, stand out from the sample with untypically low stellar luminosity. 
The location of these objects below the MS in the HR diagram (Fig.~\ref{fig:hrd}) suggests
problems in the estimation of $L_{\rm *}$ by the strong circumstellar contribution. 
Wrong values for $L_{\rm *}$ will also affect the $L_{\rm x}/L_{\rm *}$ ratio.  
In fact, V892\,Tau has the by far highest fractional X-ray luminosity of the HAeBe stars,  
$\log{(L_{\rm x}/L_{\rm *})} > -3$. The $L_{\rm x}/L_{\rm *}$ ratios given 
for this star in the previous literature were orders of magnitude lower, but based on 
the often cited value of $L_{\rm bol} = 38\,L_\odot$ which is dominated by the
contribution from circumstellar material.  

The {\em Chandra} data constrains the fractional X-ray luminosity of 
all non-detections of HAeBe stars below $L_{\rm x}/L_{\rm *} \sim 10^{-6}$. 
Their range of $L_{\rm x}$ is similar to that observed in the intermediate-mass stars of the ONC, 
but the upper limits obtained for the ONC are lower due to the exceptional sensitivity of the COUP.
At the age of Tucanae all B- and
A-type stars result undetected, and luminosities in excess of $\sim 10^{29}$\,erg/s can be excluded. 
Finally, the MS B-type sample covers an $L_{\rm x}$ range similar to the HAeBe stars (although none of them
are found at the high end of the distribution for HAeBe stars), but there may be a bias towards (strong)
X-ray emitters as a result of the selection criteria. 
%
% OUTPUT FROM    plot_specparams.pro
%

\subsection{X-rays and Activity Parameters}\label{subsect:xray_stel}

In low-mass stars the X-ray luminosity is correlated with indicators
for magnetic activity, such as rotation period, H$\alpha$ emission, and radio continuum emission. 
The rotation rate of active stars is considered an indicator for the efficiency of the stellar dynamo
\citep[e.g. ][]{Pallavicini81.1}. 
Therefore, the relation between rotation rates and X-ray luminosities can be used to check the 
magnetic origin of stellar X-rays. 
We have used $v \sin{i}$ as a proxy for rotation, because 
the rotation periods of the HAeBe stars presented in this work are not known. 
Photometric monitoring by \citet{Herbst99.1} searching for periodic variability induced by star spots 
has led to mostly negative results. Optical variability is present in HAeBe stars, but seems not to 
be cyclic. This kind of variability was termed the UX\,Ori phenomenon, and it has been ascribed to either 
variable mass accretion or variable circumstellar obscuration by orbiting dust clumps 
\citep{Natta97.1, Herbst99.1}. Therefore, it may not be surprising that we find no evidence for a 
correlation of $v \sin{i}$ with the X-ray luminosity. 
This is also consistent with earlier results from the studies by \citet{Zinnecker94.1} and 
\citet{Hamaguchi05.1}.  
In any case, it suggests that -- if anything -- only a dynamo which is independent of rotation can be at 
work. 

\begin{figure}
\begin{center}
\resizebox{9cm}{!}{\includegraphics{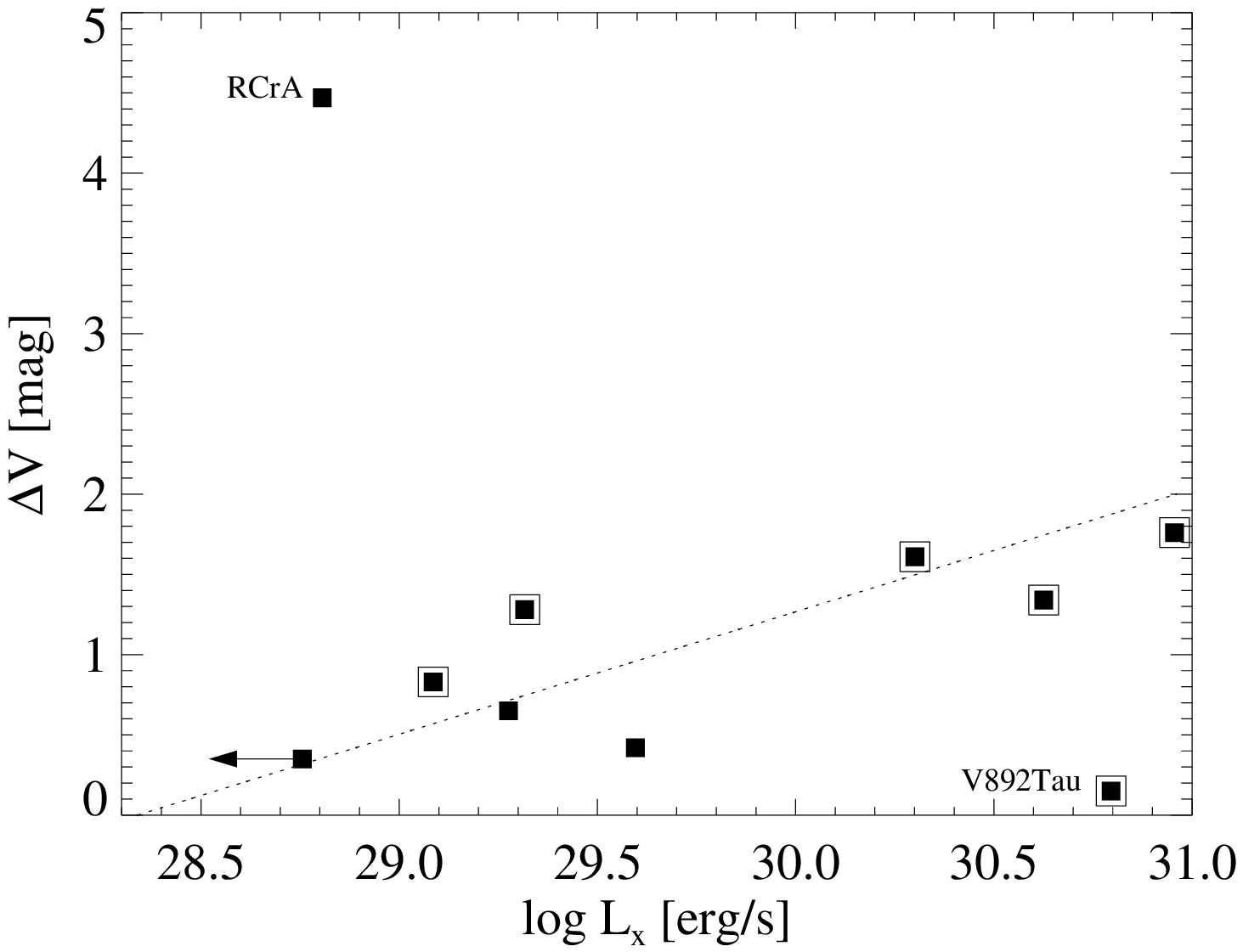}}
\caption{X-ray luminosity for HAeBe stars versus the amplitude of the $V$ band magnitude from 
\protect\citet{Herbst99.1}. The dotted line represents a linear regression fit to the data where
the two outliers (R\,CrA and V892\,Tau) are excluded.}
\label{fig:lx_dV}
\end{center}
\end{figure}

Measurements from optical monitoring are given by \citet{Herbst99.1} for $10$ HAeBe stars from the 
{\em Chandra} sample. 
With the exception of two stars, there seems to be a trend towards larger variability amplitudes
for more X-ray luminous stars (see Fig.~\ref{fig:lx_dV}). 
Note, that the $V$ band amplitude for V892\,Tau was determined from only
two measurements, while for all other stars the number of observations ranges around several hundred. 
The other outlier, R\,CrA, is an extreme case of an UXor object, displaying $V$ band variations in 
excess of $4$\,mag. 
Note, that \citet{Herbst99.1} noted a trend for optically brighter stars to show larger amplitudes. 
However, the possible connection between $\Delta V$ and $L_{\rm x}$ shown in Fig.~\ref{fig:lx_dV} can not
be caused by a relation between $L_{\rm x}$ and $L_{\rm *}$, because such a correlation is not present in the 
HAeBe sample (see Fig.~\ref{fig:lx_lbol}) where most stars span a small range in stellar 
luminosity. 
%
% OUTPUT FROM    plot_specparams.pro
%
 
The H$\alpha$ flux of late-type dwarf stars in the field is typically
a factor $3$ lower than the X-ray flux \citep{Hawley03.1}. Furthermore, all active stars 
follow a linear relation between the logarithms of their X-ray and radio luminosities \citep{Guedel93.1,
Guedel93.2}. 
However, this is not the case for the HAeBe stars. 
They are strongly overluminous in both H$\alpha$ %(Fig.~\ref{fig:lha_lx}) 
and radio, and no trend with X-ray emission is seen.

\section{Conclusions}\label{sect:conclusions}

We have presented the {\em Chandra} observations of $17$ HAeBe stars, 
with particular focus on the possible role that late-type companions play in the
explanation of the observed X-ray emission. We have compared their X-ray properties 
to a sample of MS B-type stars and B/A-type stars in the ONC observed with the same instrument,
as well as to late-type pre-MS stars. 

Based on the present status on their multiplicity, 
the X-ray emission can {\it not} be explained by known companion stars in a fraction of $\sim 35$\,\% 
of the sample. To our knowledge no systematic studies of binarity in HAeBe stars
have been published yet. It is conceivable that all X-rays come from as yet undiscovered 
late-type companions. Pursuing this line of thought  
the X-ray detection fraction of HAeBe systems in our study 
would imply a fraction of non-single stars as high as $76$\,\%. 
In practice, however, the {\em Chandra} sample
may be biased towards stars that had been detected in previous X-ray surveys, and may not be representative
for the total population of HAeBe stars, such that  
this number must be considered an upper limit to the actual binary fraction. 

Alternatively, all or some of the HAeBe stars associated with {\em Chandra} sources
may produce X-ray emission, either in coronae, star-disk magnetospheres, or winds.
A comparison of their X-ray characteristics to more evolved intermediate-mass stars is instructive, 
because several of these mechanisms predict declining activity with age. 
The detection rates 
for HAeBe stars and MS B-type stars cannot be compared directly, because of different sample selection,  
but there is a clear distinction in terms of X-ray properties. Confirming earlier results obtained
with {\em ASCA} we find that HAeBe systems have significantly hotter
X-ray plasma than MS B-stars and/or their companions, and somewhat higher X-ray luminosities. 
These differences might be a sign for fading dynamo action. 

Unfortunately, more than one mechanism is conceivable that goes along with a decrease of activity. 
One possibility of field generation in fully radiative stars is a shear dynamo, that weakens in 
the course of MS evolution because it is based on the diminuishing rotational energy of the star.
On the other hand, the X-ray luminosity and temperature of magnetically active late-type 
stars also decreases with age \citep{Stelzer00.1, Kastner03.1, Preibisch05.1}.   
Therefore, cool companions to HAeBe stars are expected to have hotter coronae than cool companions 
to MS stars, and all observed X-rays might again be traced back to potential companions. 
We have shown that the distribution of X-ray temperatures measured for the HAeBe stars is similar
to the distribution for TTS in Orion, consistent with this scenario.  

Radiatively driven winds are precluded as cause for the X-rays from HAeBe stars: 
The terminal wind speeds ($\sim 300$\,km/s) yield a maximum temperature of a few MK at most 
under the assumption of strong shocks, while the observed X-ray temperatures are higher ($\sim 20$\,MK).
For similar reasons, a magnetically confined wind, similar to the model 
introduced by \citet{Babel97.1} for Ap stars, is unlikely to apply. 
In this scenario X-rays are produced in the closed part of the magnetosphere as a consequence of shocks 
that form upon collision of the winds from the two hemispheres 
near the equatorial plane.    
The X-ray luminosities predicted by the model are consistent with the observed values, but the HAeBe 
winds are too slow to explain the observed high X-ray temperature.  
  
The absence of correlations with activity parameters (H$\alpha$ emission, radio continuum emission, 
rotation rate) 
does  not per se speak against magnetic processes on the HAeBe stars, because the contribution from magnetic
activity in the optical and radio may be masked by the stronger effects of winds. 
Another possibility is X-ray emission related to an accretion shock, similar to the case of cTTS. 
Magnetospheric accretion has been invoked to explain the shape of H$\alpha$ line profiles 
for some HAeBe stars \citep{Muzerolle04.1, Deleuil04.1}, but, again, the free-fall velocities 
($\sim 600$\,km/s) are inconsistent with the observed high X-ray temperatures. 

The incidence of magnetic fields in $\sim 10$\,\% of HAeBe stars 
is similar to the fraction of magnetic Ap/Bp stars on the MS, 
and has been taken for evidence that HAeBe stars are the progenitors of magnetic Ap/Bp stars 
\citep{Wade05.1}. 
If the X-ray emission is related in some way to these fields, 
one would expect that most of the HAeBe stars of the
{\em Chandra} sample belong to this small fraction of magnetic stars. 

To resolve the mystery of the origin of the X-ray emission from HAeBe stars 
we strongly encourage the search for magnetic fields on X-ray emitting HAeBe stars, 
systematic binarity studies of X-ray observed samples to test the companion hypothesis,  
X-ray surveys on a large and unbiased sample to better constrain detection rates and global X-ray properties, 
X-ray variability studies of HAeBe stars recognized to be single to search for magnetic activity, 
and high spectral resolution X-ray studies of individual objects to probe the density of the emitting
plasma.

\appendix

\section{Individual Systems}\label{sect:indiv}

\subsection{BD+30$^\circ$549}\label{subsect:hip16243}

The B9 star BD\,$+30^\circ549$ is the optically brightest star of the NGC\,1333 star 
forming region.

Undetected with both {\em ROSAT} and {\em ASCA} \citep{Hamaguchi05.1}, 
the first X-ray detection of this HAeBe star was presented by \citet{Getman02.1}
based on the {\em Chandra} observation; source\,83. 
Our re-analysis of the {\em Chandra} image confirms that the X-ray luminosity amounts to 
$\log{L_{\rm x}} = 29.3$\,erg/s, well below the detection limits of the previous X-ray 
observations. 
A weak, non-variable source at the position of BD\,$+30^\circ549$ 
was also seen with {\em XMM-Newton} \citep{Preibisch03.1}. 
The spectral parameters we derive are in agreement with previous determinations
by \citet{Getman02.1} and \citet{Preibisch03.1}.

\subsection{V892\,Tau}\label{subsect:v892tau}

V892\,Tau is a HAeBe star in the Taurus cloud complex. Next to the primary
of spectral type B9...A6 the system includes a T Tauri type companion at 
a separation of $\sim 4^{\prime\prime}$ \citep{Leinert97.1}.  
\citet{Smith05.1} discovered a further faint companion with estimated mass of
$1.5-2\,M_\odot$ at only $50$\,mas from the HAeBe star. 

V892\,Tau was observed by both {\em XMM-Newton} and {\em Chandra},
however only {\em Chandra} was able to resolve the wider pair. Two X-ray
sources are seen. The brighter one was identified by \citet{Giardino04.1} with
the HAeBe star. This observation was particularly intriguing because of an 
X-ray flare in the {\em Chandra} data (see Fig.~\ref{fig:acis_lcs}). After the discovery of
the further sub-arcsec companion, which remains unresolved in the X-ray data, 
the association of the flare with the HAeBe star must now be questioned. 
This companion possibly features a shallow convection zone, and 
may have $L_{\rm *}$ almost as high as the HAeBe primary. Therefore, it would yield 
a `reasonable' value for $\log{(L_{\rm x}/L_{\rm *})}$ if responsible for the X-ray emission. 

The spectrum we extract in our re-analysis of
the {\em Chandra} data indicates a $\sim 2$\,keV plasma for the brighter X-ray 
component (identified with the system of HAeBe star plus $50$\,mas companion),
absorbed by a hydrogen column density of 
$\sim 1.4 \times 10^{22}\,{\rm cm^{-2}}$, and a luminosity of $10^{30.7}$\,erg/s, in rough 
accordance with the values quoted by \citet{Giardino04.1}. Note, that we used
a slightly different distance (162\,pc versus 140\,pc). 
The emission measure of the companion, V892Tau\,NE, 
is smaller by a factor of $10$, has a soft component very similar in shape to
that of V892Tau\,A, but the hard component is weaker.

\subsection{V380\,Ori}\label{subsect:v380ori}

V380\,Ori is a visual binary of separation $0.154^{\prime\prime}$ \citep{Leinert97.1}.
Probably both components are HAeBe stars. In addition \citet{Corporon99.1} detected
the Li line in the spectrum of V380\,Ori\,A, indicating the existence of a further 
late-type companion. \citet{Testi98.1} found no evidence for additional faint IR sources
surrounding the star, and argued that the detection of such companions might be impeded
by the diffuse nebulosity seen in the $J$ and $K$ band images. 

V380\,Ori was detected as a strong X-ray source during a {\em ROSAT} pointing
\citep{Zinnecker94.1}. An X-ray spectrum obtained with {\em ASCA} showed rather
high plasma temperature \citep[$\sim 3.2$\,keV;][]{Hamaguchi05.1}. 
V380\,Ori is in the field of a $20$\,ksec long archived 
{\em Chandra} pointing whose prime target was the Herbig-Haro object HH\,1. 
Due to the small binary separation even the superior angular resolution of {\em Chandra}
is not sufficient to resolve the system, such that the origin of the single X-ray source 
remains obscure. 
The {\em Chandra} spectrum can be described reasonably well with a 1-T model of 
$kT \sim 1.6$\,keV, and column density within a factor of two of the value that corresponds
to the visual extinction. 
The X-ray luminosity ($\log{L_{\rm x}} = 31.3$\,erg/s) is at the upper end of 
the range observed in this sample. Previous X-ray satellites have measured similarly high
X-ray luminosity. %(\cite{Zinnecker94.1}, \cite{Hamaguchi05.1}). 

A longitudinal magnetic field of $0.5$\,kG was detected in spectropolarimetric
measurements \citep{Wade05.1}.

\subsection{HD\,147889}\label{subsect:hd147889}

HD\,147889 is located in the $\rho$\,Oph cloud complex. It is a known double-lined spectroscopic 
binary composed of two nearly equal B2-type stars \citep{Haffner95.1}, 
but has no known visual companion. In particular, 
a speckle search yielded a negative result \citep{Ratzka05.1}. 
It is generally assumed that HD\,147889 lies behind denser material than
other lines of sight within $\rho$\,Oph \citep[see ][ and references therein]{Snow86.1}. 

HD\,147889 was in the sample of HAeBe stars studied by \citet{Hamaguchi05.1},
but no X-ray emission was reported. The star is also undetected with {\em Chandra}
down to a detection limit of $\log{L_{\rm x}} < 28.6$\,erg/s, 
corresponding to one of the strongest constraints for the fractional X-ray luminosities 
of HAeBe stars measured so far $\log{(L_{\rm x}/L_{\rm *})} < -8.4$.

\subsection{V590\,Mon}\label{subsect:v590mon}

V590\,Mon (=LkH$\alpha\,25$) is an enigmatic object in the NGC\,2264 cluster: in the HR diagram 
it is located below the MS but it shows signs for pre-MS status,  most notably an extreme
IR excess \citep{Rydgren87.1}, leading to the speculation that this object suffers 
significant extinction arising from a disk viewed edge-on.  
From the absence of both mm- or cm-continuum emission \citet{Fuente98.1} concluded that 
this star is located in a cavity of the molecular cloud. 
The speckle search for binaries among HAeBe stars by
\citet{Leinert97.1} resulted negative for this star. 

The NGC\,2264 cluster was observed for 50\,ksec with {\em Chandra} \citep{Ramirez04.1}. 
We find an upper limit of $\log{L_{\rm x}} < 28.8$ for V590\,Mon, consistent with the 
value cited by \citet{Ramirez04.1}. Its non-detection is in accordance 
with earlier (less sensitive) X-ray observations. An X-ray source is identified $\sim 10.6$
North of V590\,Mon in the small portion of the image we analysed,
%% at position: 06:40:44.583 , +09:48:12.76
corresponding to source no.103 of \citet{Ramirez04.1}, a K7 star.

\subsection{Z\,CMa}\label{subsect:zcma}

The perception of Z\,CMa, originally classified as HAeBe star \citep{Herbig60.1}, has
dramatically changed in the course of the years, as more and more observational details
were revealed. 
%Vastly different values for the stellar parameters are found in the literature, with the extremes
%placing it above the birthline in the HR diagram. 

Z\,CMa is a young F-type star with 
FU\,Ori like characteristics, a mass of $\sim3\,M_\odot$, bolometric luminosity of 
$28\,L_\odot$, and an age of $\sim 3 \times 10^5$\,yrs,  
just below the birthline in the HR diagram \citep{Hartmann89.1}. 
A powerful IR source is found at only $0.1^{\prime\prime}$ \citep{Koresko91.1}. 
Assuming that this object is coeval with the FU\,Or star, \citet{vandenAncker04.1} showed that it
can be modeled as a B0 type star on the birthline 
with a luminosity of $\log{(L_*/L_\odot)} > 5$ and a mass of $38\,M_\odot$.
Therefore, this is the actual primary of the system. 
In fact, in traditional models (see Fig.~\ref{fig:hrd}) the star is placed above 
the birthline. However, calculations that include the effect of dust on the evolution of a massive
star raise the birthline \citep{Yorke02.1}, such that the high observed luminosity can be accomodated. 

The system is characterized by occasional outbursts in optical brightness.
Although the optical light is usually dominated by the FU\,Or star, 
the spectrum during outburst is dominated by the IR companion, which drives a jet and
bipolar outflow \citep{Garcia99.1}. The origin of the outbursts is
not well understood. They may be related to changes in the magnetic field configuration that entail
changes in the outflow \citep{vandenAncker04.1}. 

With {\em Chandra} we find two X-ray sources in the vicinity of Z\,CMa, one on the position 
of the unresolved binary and another unidentified and very faint one $\sim 8.6^{\prime\prime}$ 
to the south. 
If the stellar luminosity of Z\,CMa is indeed as high as assumed throughout this paper 
\citep[$\log{(L_*/L_\odot)} = 5.15$; ][]{Acke04.1},  
the upper limit for $\log{(L_{\rm x}/L_{\rm *})}$ ($\sim -8.4$) comes out very low  
despite the fact that the X-ray luminosity is among the highest of the HAeBe stars.

\subsection{HD\,97300}\label{subsect:hd97300}

HD\,97300 is located in the Cha\,I star forming region. 
Mid-IR images show that HD\,97300 is surrounded by a dust ring possibly created by interaction 
of the star with circumstellar matter, and a secondary peak of the ISOCAM image suggests the presence
of an IR companion $\sim 3^{\prime\prime}$ north of the primary which is detected only above $10\,\mu$m
\citep{Siebenmorgen98.1}.

HD\,97300 was detected with {\em ROSAT} \citep[see ][]{Hamaguchi05.1} at $\log{L_{\rm x}}=29.0$\,erg/s. 
\citet{Feigelson04.1} discussed the {\em Chandra} image of Cha\,I North, where HD\,97300 was
identified as their source\,44. 
In our restricted {\em Chandra} image the only X-ray source, brighter by $\sim 1$\,dex than deduced from
the {\em ROSAT} data, is clearly associated with HD\,97300 itself. 
The decaying phase of a flare is seen at the beginning of the observation, but the flare is not
strong enough to make a significant difference between the quiescent and the average count rate.
There is no evidence for X-ray emission from the IR companion.

\subsection{HD\,100546}\label{subsect:hd100546}

HD\,100546 is a B9.5\,Ve HAeBe star \citep{Hu89.1}, and 
with an age of $> 10$\,Myrs one of the most evolved objects of this class
\citep{vandenAncker97.1}.  
Evidence for its rather large age comes from the 
$\lambda$\,Boo like abundance anomalies 
that hint at gas-dust separation and ensuing metal depletion of the accreted
material, thought to take place in the late pre-MS phase \citep{Acke04.2}. 
\citet{Augereau01.1} and \citet{Shatsky02.1} 
reported the detection of four faint IR objects within 
$\sim 10^{\prime\prime}$ of HD\,100546 
\citep[see also ][ for a cartoon on the environment of HD\,100546]{Grady01.1}. 
Based on the near-IR photometry all four companions were 
classified as `optical', i.e. no true physical companions.
This conclusion is in agreement with the non-detection of these objects
with {\em Chandra}. Their limiting X-ray luminosity is 
$\log{L_{\rm x}} = 27.9$\,erg/s if assumed to be at the same distance as 
HD\,100546.

HD\,100546 is surrounded by a circumstellar dust disk with cometary-like
chemical composition. 
The {\em Chandra} observation of HD\,100546 was discussed by F03. 
We found no previous reports on X-ray emission from this star.
In our re-analysis of the {\em Chandra} data we detect it at 
$\log{L_{\rm x}} = 28.9$\,erg/s, 
discrepant with the value given by F03 by $0.4$\,dex. Note, that the plasma
temperature we assume in deriving the X-ray luminosity is smaller 
($1$\,keV versus $2.5$\,keV of F03).
As the absorption is similar (fixed to $A_{\rm V} = 0.06$ in our case and neglected
by F03) the difference in $L_{\rm x}$ remains unexplained.

\subsection{HD\,104237}\label{subsect:hd104237}

HD\,104237 is the optically brightest Herbig star. It is located 
in the Chamaeleon star forming complex, but outside the boundaries of the 
molecular clouds. Although showing signatures typical for HAeBe stars (emission lines,
IR excess from a dust shell) HD\,104237 is not surrounded by
a reflection nebula \citep{Hu89.1} that would indicate association with
a star forming cloud.

Furthermore, from an analysis of radial velocities and the presence of lithium
absorption in the spectrum of HD\,104237 \citet{Boehm04.1} concluded that it 
is a spectroscopic binary. 
An {\em IUE} observation of HD\,104237 has shown evidence for
hot plasma \citep{Hu91.1}. The star is listed as a RASS 
X-ray source by \citet{Alcala95.1}, and was
also detected in a dedicated observation by {\em ASCA} 
\citep{Skinner96.1, Hamaguchi05.1}. 
\citet{Feigelson03.1} resolved the X-ray emission from this object with
two {\em Chandra} snapshots into five individual sources 
that likely form a young stellar group. 
No close visual companions were known prior to this study, but optical/IR identifications
for the new X-ray sources have meanwhile been provided \citep{Grady04.1}. 

In our reanalysis of the {\em Chandra} data we find only three sources, identified
with HD\,104237\,A, HD\,104237-5, and HD\,104237-6. The brightest one is the HAeBe star. 
As discussed in the text, HD\,104237-2 is probably missed because of its small separation 
from HD\,104237\,A.

\subsection{HD\,141569}\label{subsect:hd141569}

The evolutionary state of HD\,141569 is somewhat ambiguous. It exhibits 
characteristics of both HAeBe stars (emission lines, association with a reflection
nebula) and MS stars (low ratio of infrared to stellar luminosity). 
\citet{vandenAncker98.1} conclude that HD\,141569 may be a transition object,
consistent with its position in the HR diagram very close to the ZAMS.
The $4^{\prime\prime}$ wide disk was imaged in scattered light 
\citep{Weinberger99.1, Augereau99.1}. 
These images were sensitive to the detection of planets down to 
$3\,M_{\rm Jup}$, but did  not detect any such object, thus the existence of 
very close-in companions is unlikely. 
In a search for spectroscopic binaries among HAeBe stars HD\,141569 provided
a negative result \citep{Corporon99.1}. 
 
Two potential companions within $< 10^{\prime\prime}$ of HD\,141569 have
first been identified by \citet{Rossiter43.1}. Recently the position and
IR photometry of these objects were discussed by \citet{Weinberger00.1}.
These have shown that the system forms a physical triple: 
Astrometry confirms that both companions are bound,
and the presence of lithium absorption in their spectrum 
establishes their pre-MS nature. 

The {\em Chandra} observation of HD\,141569 was presented by \citet{Feigelson03.1},
who reported a detection for HD\,141569\,A but no X-ray emission from the two visual
companions. These results were picked up by \citet{Skinner04.1}.
In our re-analysis of the same data we find another interpretation for the assignment
of counterparts to the X-ray sources:
As shown in Fig.~\ref{fig:acis_images_haebe} the two X-ray sources can clarly be identified 
with the two companions, while the HAeBe star is X-ray dark down to the detection
limit, with $L_{\rm x} < 10^{28}$\,erg/s.

\subsection{HD\,150193}\label{subsect:hd150193}

This HAeBe star, also known as MWC\,863, is located near the $\rho$\,Oph cloud, 
but isolated. 
A visual companion was first discovered by \citet{Reipurth93.1}. This
object shows T Tauri star characteristics \citep{Bouvier01.1} and
is coeval with the primary \citep{Fukagawa03.1}.
Later HD\,150193\,A turned out to be a spectroscopic binary with an additional cool 
companion identified by its Li absorption \citep{Corporon99.1}. 
The primary in this system is surrounded by a circumstellar disk. 
No disk was detected around the visual companion. 

Based on the {\em Chandra} image \citet{Feigelson03.1} considered the companion
undetected, and claimed a new companion `C' identified by means of its X-ray emission. 
In our re-analysis of the {\em Chandra} data we detect a single, elongated X-ray source midway between the HAeBe 
star and its known companion located south-west of the primary. Visually we identify
two marginally separated sources, one of which is the `companion C' of F03. 
However, our cross-correlation with the optical/IR positions
suggests that this weak X-ray source to the NE of the brighter source should be
identified with the HAeBe star, while the brighter source itself coincides with the 
T Tauri companion \citep[see our Fig.~\ref{fig:acis_images_haebe} and ][]{Skinner04.1}.

\subsection{HD\,152404}\label{subsect:hd152404}

Strictly speaking HD\,152404 is not a HAeBe star, because its spectral type
is F5. The Upper\,Sco member \citep{deZeeuw99.1} 
is an eccentric double-lined spectroscopic binary system 
composed of two nearly equal mass stars \citep{Andersen89.1}. 
The profiles of the Balmer emission lines provide evidence for both accretion and
outflow, and the SED can be modeled by a circumbinary disk with a large inner gap 
\citep{Alencar03.1}

At the position of HD\,152404 the {\em Chandra} image presents one faint 
X-ray source \citep[see Fig.~\ref{fig:acis_images_haebe}), also reported by ][]{Feigelson03.1}.

\subsection{HD\,163296}\label{subsect:hd163296}

HD\,163296 is a well-studied isolated nearby HAeBe star of intermediate age.
It has a dust disk, %of $\sim 3.7^{\prime\prime}$ radius, 
and is associated with a series
of HH objects \citep{Grady00.1}. The rich emission line spectrum of HD\,163926
was studied by \citet{Deleuil05.1}, revealing P\,Cyg features indicative of a wind
and evidence for a chromosphere (heated region). The line profiles of highly ionized species
can be explained by magnetospheric accretion or by magnetospherically confined wind shocks.  

The {\em Chandra} observation of HD\,163926 was discussed by \citet{Swartz05.1},
who derived spectral parameters compatible with our results. They speculated that
this star may represent a case of X-ray emission from an accretion
shock because of the untypically low temperature. 
Another possibility is that the soft X-rays arise from shocks at the interface
of the accretion disk with the optical jet, as in the cases of L1551\,IRS\,5 \citep{Bally03.1}
and the Beehive Proplyd \citep{Kastner05.1}.

\subsection{MWC\,297}\label{subsect:mwc297}

MWC\,297 is one of the hottest and most absorbed HAeBe stars (spectral type O9\,e;
$A_{\rm V} = 8$\,mag). \citet{Drew97.1} studied this object in detail, and
provided an updated distance estimate. Recently a circumstellar disk was 
resolved \citep{Eisner04.1}.
\citet{Vink05.1} have identified a faint object at $3.4^{\prime\prime}$
from MWC\,297 in coronographic adaptive optics $H$-band images. 

In light of the discovery of a close companion, the report of a
large X-ray flare from MWC\,297 in low-spatial resolution {\em ASCA} data 
by \citet{Hamaguchi05.1} must be questioned. 
If it truely occurred on MWC\,297 this would be one of
only two X-ray flares ascribed to HAeBe stars so far. However, \citet{Damiani06.1}
have shown from their analysis of {\em Chandra} data that there are no less than 
$9$ X-ray sources in the {\em ASCA} point-spread function, and MWC\,297 is not
even the brightest of these. We have analysed the same {\em Chandra} data and
found two sources in the small ($100 \times 100$\,pixels) image. These two X-ray sources coincide with
MWC\,297 and its adaptive optics companion. Both sources are too faint for spectral analysis. 
But the X-ray luminosity for the MWC\,297 source ($\log{L_{\rm x}} = 29.3$\,erg/s) 
is magnitudes below the estimate from {\em ASCA} ($\log{L_{\rm x}} \sim 31...32$\,erg/s).

\subsection{HD\,176386}\label{subsect:hd176386}

HD\,176386 shows far-IR excess detected with IRAS indicating emission from a dust shell.
Its mid-IR spectrum is dominated by extended emission \citep{Prusti94.1}
and shows two peaks, the stronger of which is associated with the position
of the star \citep{Siebenmorgen00.1}.  
Although close to the ZAMS in the HR diagram, the presence of C\,IV absorption lines in 
IUE spectra of HD\,176386 with wings extending out to velocities of several hundreds of km 
was interpreted as evidence for accretion of circumstellar gas onto the star \citep{Grady93.1}.

HD\,176386 shares proper motion with TY\,CrA, from which it is separated by $\sim 55^{\prime\prime}$, 
and the two stars may be physically associated with each other \citep{Teixeira00.1}.
From the {\em Hipparcos} data base (REF) we find that HD\,176386 is a binary with $4.1^{\prime\prime}$
separation. The secondary is $\sim 5$\,mag fainter. 

\citet{Hamaguchi05.1} reported an X-ray detection of HD\,176386 with {\em ASCA}, but the object was
only marginally resolved from TY\,CrA. The {\em Chandra} data presented here not only confirms the detection, 
but demonstrates also that the X-ray source is associated with the companion. The HAeBe star itself
is X-ray dark down to our detection limit of $\log{L_{\rm x}} < 28.6$\,erg/s.

\subsection{TY\,CrA}\label{subsect:tycra}

TY\,CrA forms a multiple system. The central eclipsing binary \citep{Kardopolov81.1},
a spectroscopic companion with $M \sin^3{i} = 2.4 \pm 0.5\,M_\odot$ \citep{Casey95.1},
and a visual $\sim$M4-type companion at $\sim0.3^{\prime\prime}$ separation \citep{Chauvin03.1}.
The central binary system is composed of the Herbig Be primary and a $\sim 1.6\,M\odot$ secondary star  
in a circular orbit with period of $\sim 2.89$\,d \citep{Casey93.1}.
The lack of near-IR emission and the small extinction of $3$\,mag
does not favor the presence of an optically thick disk \citep{Casey93.1}.
 
X-ray emission from TY\,CrA has been observed with {\em ROSAT}, {\em Einstein} and {\em ASCA}
\citep[][]{Zinnecker94.1, Damiani94.1, Hamaguchi05.1}.
The X-ray properties of all these observations are in good agreement with our {\em Chandra} measurements.
% ASCA: $N_{\rm H}\sim 4 \times 10^{21}\,{\rm cm^{-2}}$, $kT \sim 1-1.5$\,keV, $\log{L_{\rm x}} \sim 30.4-30.5$\,erg/s. 
It is one of the X-ray brightest HAeBe stars ($\log{L_{\rm x}} \sim 30.9$\,erg/s). 
TY\,CrA has shown a flare during the {\em Chandra} observation, during which it has roughly doubled its count rate 
(Fig.~\ref{fig:acis_lcs}). The individual components of the quadruple 
can not be resolved, such that the only X-ray source
may in principle be composed of contributions from all four stars.

\subsection{R\,CrA}\label{subsect:rcra}

R\,CrA is a presumably single HAeBe star \citep{Leinert97.1} 
surrounded by a reflection nebula that scatters its variable light. 
The molecular cloud surrounding R\,CrA contains a number of embedded
IR objects, that might be the driving source for some nearby HH outflows 
\citep{Anderson97.1}.

While R\,CrA remained undetected with {\em ROSAT} \citep{Zinnecker94.1} 
and {\em ASCA} \citep{Hamaguchi05.1}, 
a recent X-ray detection with {\em Chandra} was discussed by \citet{Skinner04.1}
and \citet{Forbrich06.1}. 
In our re-analysis of the {\em Chandra} image we derive a luminosity 
of $\log{L_{\rm x}} = 28.8$\,erg/s, lower by a factor of two from the value cited
by \citet{Skinner04.1}. The difference can partly be attributed to the different
distance estimate used. The X-ray spectrum is dominated by hard emission, and is similar
to a nearby class\,I protostar IRS\,9.

\begin{acknowledgements}
BS wishes to thank E. Flaccomio for stimulating discussions. 
This research has made use of the SIMBAD database, operated at CDS, Strasbourg, France,
and the {\em Hipparcos} catalogue accessed through the VizieR data base. 
\end{acknowledgements}

\end{document}